\documentclass[sigconf]{acmart}

\AtBeginDocument{%
  }

\copyrightyear{2026}
\acmYear{2026}
\setcopyright{cc}
\setcctype{by}
\acmConference[KDD '26]{Proceedings of the 32nd ACM SIGKDD Conference on Knowledge Discovery and Data Mining V.2}{August 09--13, 2026}{Jeju Island, Republic of Korea}
\acmBooktitle{Proceedings of the 32nd ACM SIGKDD Conference on Knowledge Discovery and Data Mining V.2 (KDD '26), August 09--13, 2026, Jeju Island, Republic of Korea}
\acmISBN{979-8-4007-2259-2/2026/08}
\acmDOI{10.1145/3770855.3817454}

\usepackage{amsmath,amsfonts}
\usepackage{amsthm}

\usepackage[ruled,vlined,linesnumbered]{algorithm2e}
\usepackage{algpseudocode}
\usepackage{graphicx}
\usepackage{textcomp}
\usepackage{xcolor}
\definecolor{brown}{RGB}{139,69,19}
\newcommand{\bchange}[1]{\textcolor{black}{#1}}

\usepackage{enumitem}
\usepackage{multirow}
\usepackage{booktabs}
\usepackage{tcolorbox}
\tcbuselibrary{breakable}

\usepackage{arydshln}
\usepackage{subcaption}
\usepackage{balance}

\begin{document}

\title{FDABench: A Benchmark for Data Agents on Analytical Queries over Heterogeneous Data}

\author{Ziting Wang}
\affiliation{%
  \institution{Nanyang Technological University}
  \country{Singapore}
}
\email{ziting001@e.ntu.edu.sg}

\author{Shize Zhang}
\affiliation{%
  \institution{National University of Singapore}
  \country{Singapore}
}
\email{shize.zhang@u.nus.edu}

\author{Haitao Yuan}
\affiliation{%
  \institution{Nanyang Technological University}
  \country{Singapore}
}
\email{haitao.yuan@ntu.edu.sg}

\author{Jinwei Zhu}
\affiliation{%
  \institution{Huawei Technologies Ltd.}
  \city{Shanghai}
  \country{China}
}
\email{zhujinwei@huawei.com}

\author{Wei Dong}
\affiliation{%
  \institution{Nanyang Technological University}
  \country{Singapore}
}
\email{wei_dong@ntu.edu.sg}

\author{Gao Cong}
\affiliation{%
  \institution{Nanyang Technological University}
  \country{Singapore}
}
\email{gaocong@ntu.edu.sg}

\renewcommand{\shortauthors}{Ziting Wang et al.}

\begin{abstract}
\bchange{The growing demand for data-driven decision-making has created an urgent need for data agents that can reason over heterogeneous data (databases, documents, web content, images, videos, and audio) to answer complex analytical queries.
However, evaluating such agents remains challenging: existing benchmarks often focus on isolated agent capabilities or limited data modalities, lacking comprehensive coverage of heterogeneous data and rigorous evaluation across diverse data agent architectures. To address these challenges, we present FDABench, a benchmark for evaluating data agents' reasoning ability over heterogeneous data in analytical scenarios. Our contributions are threefold: (1) A comprehensive benchmark of 2,007 tasks spanning six data modalities with a unified, multi-granularity evaluation framework. (2) We design \textit{PUDDING}, an agentic dataset construction framework that leverages LLM generation with iterative expert validation for reliable and scalable benchmark construction. (3) Extensive experiments across diverse data agent architectures, including general analytical agents, semantic operator frameworks, and RAG-based methods, revealing key insights and guidelines for future data agent development.
Our data and source code are released at \url{https://github.com/fdabench/FDAbench}.}
\end{abstract}



\maketitle

\section{Introduction}
\label{sec:intro}
\begin{figure}
  \centering
  \includegraphics[width=0.9\linewidth]{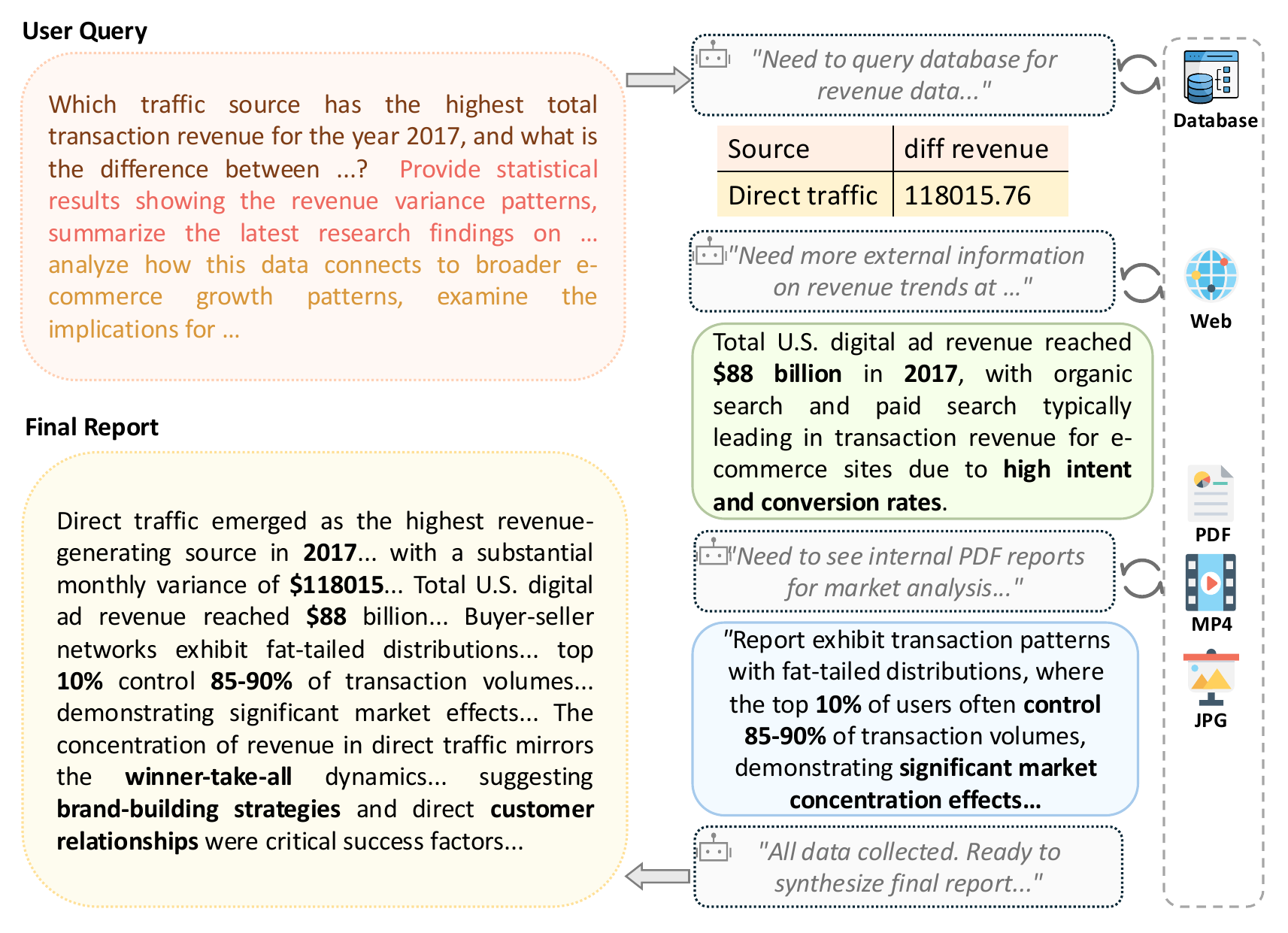}
  \vspace{-0.15in}
  \caption{Example of Analytical Query with Data Agent.}
  \vspace{-0.1in}
  \label{fig:example}
\end{figure}

\bchange{Modern data analytics increasingly requires synthesizing insights from heterogeneous data sources (i.e., databases, documents, web content, images, videos, and audio) to answer complex analytical queries~\cite{Databricks,DBLP:journals/corr/dagent,sun2025agenticdata, DBLP:journals/corr/taiji}. While LLMs have advanced tabular question answering~\cite{DBLP:journals/pvldb/AutoTQA,DBLP:journals/pacmmod/tablegpt,chat2data}, they still struggle with multi-source reasoning and precise retrieval due to hallucination~\cite{DBLP:conf/iclr/spider2,DBLP:conf/nips/agentboard,hall}. \textit{Data agents} address this challenge by positioning LLMs as coordinators that orchestrate specialized tools (e.g., SQL execution, document retrieval, web search) to decompose complex queries, synthesize results across diverse sources, and produce analytical reports~\cite{shankar2024docetl, patel2024semanticoperators, DBLP:journals/corr/dagent, DBLP:journals/corr/taiji, sun2025agenticdata, wang2025aop, Databricks}, as shown in Figure~\ref{fig:example}.}

\begin{table*}[t]
\centering
\caption{\bchange{Comparison of data-agent-related benchmarks in data modality, task focus, and evaluation supervision.}}
\label{tab:benchmark-comparison}
\renewcommand{\arraystretch}{0.95}
\setlength{\tabcolsep}{1.2pt}
\footnotesize
\begin{tabular}{lc|c|cccccc|cc|cccc}
\hline
& & & \multicolumn{6}{c|}{\bchange{\textbf{Data Modality}}} & \multicolumn{2}{c|}{\textbf{Task}} & \multicolumn{4}{c}{\bchange{\textbf{Evaluation Supervision}}} \\
\cline{4-15}
\textbf{Benchmark} & \textbf{\#Tasks} & \textbf{Focus} &
\textbf{DB} & \textbf{Doc} & \textbf{Web} & \textbf{Img} & \textbf{Vid} & \textbf{Aud} &
\textbf{Output} & \textbf{Open} &
\textbf{Eval.} & \textbf{Rubric} & \textbf{Int.} & \textbf{Trace} \\
\hline
Spider~\cite{DBLP:conf/emnlp/spider1} & 10,181 & Text2SQL  &
$\checkmark$ & $\times$ & $\times$ & $\times$ & $\times$ & $\times$ &
SQL & $\times$ &
Exec./EM & $\times$ & $\times$ & $\times$ \\
Spider 2.0~\cite{DBLP:conf/iclr/spider2} & 632 & Enterprise Text2SQL &
$\checkmark$ & $\checkmark$ & $\times$ & $\times$ & $\times$ & $\times$ &
SQL/Result & $\times$ &
Exec. & $\times$ & $\times$ & $\times$ \\
BIRD~\cite{bird} & 12,751 & Text2SQL &
$\checkmark$ & $\triangle$ & $\times$ & $\times$ & $\times$ & $\times$ &
SQL & $\times$ &
Exec. & $\times$ & $\times$ & $\times$ \\
\hline
GAIA~\cite{DBLP:conf/iclr/gaia} & 466 & Multimodal QA w/ tools &
$\times$ & $\checkmark$ & $\checkmark$ & $\checkmark$ & $\times$ & $\times$ &
Answer & $\times$ &
Obj. & $\times$ & $\times$ & $\times$ \\
WebArena~\cite{DBLP:conf/iclr/ZhouX0ZLSCOBF0N24} & 812 & Web navigation &
$\times$ & $\checkmark$ & $\checkmark$ & $\triangle$ & $\times$ & $\times$ &
Actions & $\checkmark$ &
Exec. & $\times$ & $\times$ & $\times$ \\
MINT~\cite{DBLP:conf/iclr/00020LCYPJ24} & 586 & Multi-turn tool use &
$\times$ & $\times$ & $\times$ & $\times$ & $\times$ & $\times$ &
Answer & $\times$ &
Obj./Exec. & $\times$ & $\times$ & $\times$ \\
\hline
DSBench~\cite{DBLP:conf/iclr/DSBench} & 540 & Notebook data analysis &
$\checkmark$ & $\triangle$ & $\times$ & $\checkmark$ & $\times$ & $\times$ &
Ans./Subm. & $\checkmark$ &
Obj./RPG & $\times$ & $\times$ & $\times$ \\
DA-Code~\cite{huang2024dacode} & 500 & Data analysis codegen &
$\checkmark$ & $\checkmark$ & $\times$ & $\times$ & $\times$ & $\times$ &
Code & $\checkmark$ &
Exec./Obj. & $\times$ & $\times$ & $\times$ \\
BLADE~\cite{gu2024blade} & 188 & Data science w/ notebooks &
$\checkmark$ & $\times$ & $\times$ & $\times$ & $\times$ & $\times$ &
Report/Dec. & $\checkmark$ &
F1/Dec. & $\times$ & $\checkmark$ & $\times$ \\
DS-1000~\cite{Lai2022DS1000} & 1,000 & Data science codegen &
$\triangle$ & $\times$ & $\times$ & $\times$ & $\times$ & $\times$ &
Code & $\checkmark$ &
Exec. & $\times$ & $\times$ & $\times$ \\
\hline
DABStep~\cite{DBLP:journals/corr/dabstep} & 450 & Multi-step tabular analysis &
$\checkmark$ & $\checkmark$ & $\times$ & $\times$ & $\times$ & $\times$ &
Answer & $\times$ &
Obj. & $\times$ & $\times$ & $\times$ \\
KramaBench~\cite{DBLP:journals/corr/abs-2506-06541} & 104 & Data-to-insight pipelines &
$\checkmark$ & $\checkmark$ & $\times$ & $\times$ & $\times$ & $\times$ &
Scripts & $\checkmark$ &
Exec./Func. Cov. & $\triangle$ & $\checkmark$ & $\times$ \\
DAComp~\cite{lei2025dacomp} & 210 & Data intelligence &
$\checkmark$ & $\checkmark$ & $\times$ & $\times$ & $\times$ & $\times$ &
Rep./Pipe. & $\checkmark$ &
Exec./LLM Jud. & $\checkmark$ & $\checkmark$ & $\times$ \\
\hline
\textbf{FDABench} & \textbf{2,007} & \textbf{Heterogeneous data analytics} &
$\checkmark$ & $\checkmark$ & $\checkmark$ & $\checkmark$ & $\checkmark$ & $\checkmark$ &
\textbf{Ans./Rep.} & $\checkmark$ &
\textbf{Exec./LLM Jud./DAG} & $\checkmark$ & $\checkmark$ & $\checkmark$ \\
\hline
\end{tabular}
\\
\textcolor{black}{
\raggedright\scriptsize
\textit{Note: DB = relational databases and tabular files (CSV/Excel/Parquet); Doc = unstructured documents (PDF/TXT/markdown). Exec.\ = execution match, EM = exact matching, Obj.\ = objective match, LLM Jud.\ = LLM-as-judge, RPG = Relative Performance Gap, Func.\ Cov.\ = functionality coverage, Dec.\ = decision accuracy. Int.\ = intermediate; Trace = tool-call DAG. $\triangle$ = indirect/limited support.}
}
\end{table*}

\bchange{However, existing benchmarks do not fully address evaluating data agents over heterogeneous data, as shown in Table~\ref{tab:benchmark-comparison}.
\textit{Structured data benchmarks}~\cite{DBLP:conf/emnlp/spider1, DBLP:conf/iclr/spider2, bird} are restricted to relational database tasks such as Text-to-SQL, without considering unstructured data sources.
\textit{Unstructured data benchmarks}~\cite{DBLP:conf/nips/CRAG, DBLP:journals/corr/RAGbench, DBLP:conf/nips/marco, hotpotqa} focus on retrieval and answer synthesis over text corpora like RAG scenarios, but are limited to textual retrieval and do not evaluate analytical reasoning across heterogeneous data sources.
\textit{Heterogeneous data benchmarks}~\cite{DBLP:conf/iclr/DSBench, huang2024dacode, gu2024blade, Lai2022DS1000} evaluate code generation for data analysis tasks, but focus on fixed-output scenarios and do not comprehensively evaluate data agent reasoning ability over open-ended analytical tasks. \textit{General agent benchmarks}~\cite{DBLP:conf/iclr/gaia, DBLP:conf/iclr/ZhouX0ZLSCOBF0N24, DBLP:conf/iclr/00020LCYPJ24} target web browsing or multi-turn tool interaction, but do not focus on analytical reasoning over data.
Therefore, some concurrent work, such as DABStep~\cite{DBLP:journals/corr/dabstep}, KramaBench~\cite{DBLP:journals/corr/abs-2506-06541}, and DAComp~\cite{lei2025dacomp}, extends data agent evaluation beyond single-step query answering to multi-step analytical pipelines and complex data workflows. 
However, two critical gaps still remain: (1) existing benchmarks cover at most three data modalities, yet real-world data analytical tasks involve far richer modalities~\cite{DBLP:journals/corr/abs-2401-03568};
and (2) current data agent benchmarks lack reliable reasoning trace assessment, often relying on coarse-grained LLM judge which may introduce model bias.}

\bchange{To address these gaps, we propose FDABench, a benchmark for evaluating data agents' reasoning ability over heterogeneous data in analytical scenarios.} \textcolor{black}{We identify three fundamental research questions that must be addressed:}

\noindent \textcolor{black}{\textbf{RQ1: What aspects should a data agent benchmark cover for analytical query tasks?} \bchange{The analytical query tasks involve some key factors: heterogeneous data, multi-step tool orchestration, and open-ended outputs. Therefore, we identify four key dimensions for the agent evaluation: data heterogeneity, task complexity, output diversity, and workflow variability.}}

\noindent \textcolor{black}{\textbf{RQ2: How to construct reliable test cases at scale?} \bchange{Traditional manual annotation cannot scale to data agent scenarios that require multi-step reasoning and iterative tool invocation over heterogeneous data. This necessitates data agents benchmarking construction frameworks that balance efficiency and reliability.}}

\noindent \textcolor{black}{\textbf{RQ3: How to fairly evaluate across diverse architectures?} \bchange{Current data agent implementations span multiple technical paradigms, including semantic operator frameworks~\cite{palimpzestCIDR,patel2024semanticoperators}, document processing pipelines~\cite{shankar2024docetl}, and LLM-orchestrated tool chains. Each system employs different input formats, internal workflows, and output structures. FDABench addresses this through an evaluation framework supporting different data agent workflow patterns, and three task types (single-choice, multiple-choice, and report) to handle diverse output formats.}}

\noindent \textcolor{black}{\noindent \textbf{Contributions.}
Our contributions are summarized as follows:}

\noindent $\bullet$~\textcolor{black}{\textbf{Comprehensive Benchmark with Multi-Granularity Evaluation:} \bchange{We develop FDABench, a comprehensive benchmark with 2,007 tasks covering six data modalities and three task types (single-choice, multiple-choice, and report), with a multi-granularity evaluation framework that supports both structured answers and open-ended reports. (\textbf{Addressing RQ1})}}

\noindent  $\bullet$~\textcolor{black}{\textbf{Efficient and Reliable Test Case Construction:} We design \textit{PUDDING}, an agentic dataset construction framework that \textcolor{black}{performs tree-structured exploration with per-branch self-reflection} and reliable expert validation. 
The framework achieves 48.6\% acceptance rate. (\textbf{Addressing RQ2})}

\noindent  $\bullet$~\textcolor{black}{\textbf{Unified Evaluation Framework:} \bchange{We develop a cross-system framework supporting four data agent workflows: planning, tool-use, reflection, and multi-agent.
The framework provides standardized interfaces and three evaluation layers (choice correctness, rubric-based report scoring, and DAG-based reasoning trace and tool metrics), enabling fair comparisons across diverse data agent systems. (\textbf{Addressing RQ3})
}}

\noindent  $\bullet$~\textcolor{black}{\textbf{Extensive Experiments and Insights:} \bchange{We evaluate 12 foundation models across 4 data agent workflow patterns, and benchmark against existing data agent systems, including 6 general analytical agents, 3 semantic operator frameworks, and 4 RAG-based methods.
We analyze root causes of performance variations and provide actionable insights for future data agent design.}}

\vspace{-15pt}

\section{{RELATED WORK\label{sec:related}}}

As shown in Table~\ref{tab:benchmark-comparison}, we compare FDABench with existing benchmarks and provide detailed discussions below.

\noindent \textbf{Structured Data Benchmarks.}
\bchange{Data agents processing structured data in relational databases typically employ Text-to-SQL~\cite{DBLP:journals/pvldb/text2sqlsurvery, DBLP:journals/vldb/text2sqlsurvery2,DBLP:journals/pvldb/text2sqlsurvery3,DBLP:conf/sigmod/GkiniBKI21,DBLP:journals/pvldb/LLMdatalake,NL2SQL,DBLP:journals/pacmmod/LiZLFZZWP0024,DBLP:journals/pacmmod/GuF00JM023}. Spider~\cite{DBLP:conf/emnlp/spider1} focuses on cross-domain SQL generation with 10,181 questions across 200 databases. Spider 2.0~\cite{DBLP:conf/iclr/spider2} targets enterprise-level SQL generation requiring multi-step queries across massive databases and multiple SQL dialects. BIRD~\cite{bird} addresses dirty data processing and SQL efficiency optimization. These benchmarks are restricted to relational databases and SQL generation, without considering complex heterogeneous data.}

\noindent \textbf{Unstructured Data Benchmarks.}
\bchange{CRAG~\cite{DBLP:conf/nips/CRAG} spans five domains for simulating web and knowledge graph search, while RAGBench~\cite{DBLP:journals/corr/RAGbench} evaluates across five industry domains with technical documentation. MS MARCO~\cite{DBLP:conf/nips/marco} is derived from Bing~\cite{bing} query logs, and HotpotQA~\cite{hotpotqa} necessitates multi-hop reasoning across Wikipedia articles. These benchmarks mainly focus on retrieval quality and answer synthesis over unstructured corpora, but do not comprehensively cover heterogeneous data modalities.}

\begin{figure*}
  \centering
  \includegraphics[width=0.79\linewidth]{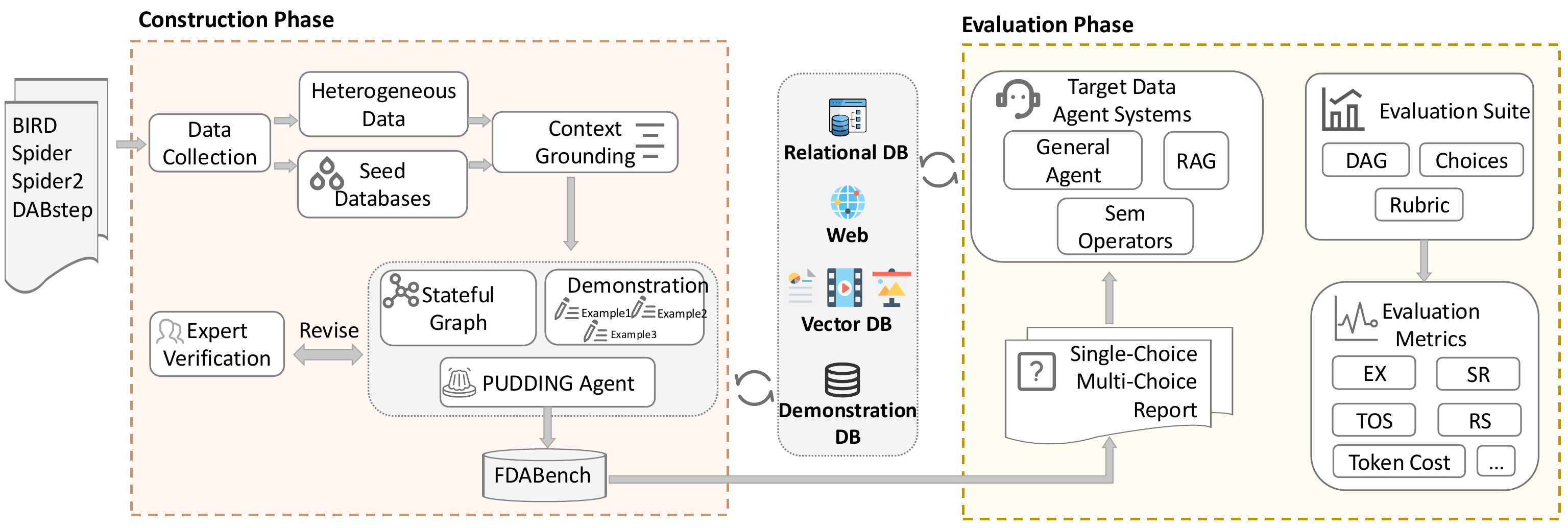}
   \vspace{-0.15in}
\caption{The Architecture Overview of FDABench.}
  \label{fig:benchmark_overview}
  \vspace{-0.1in}
\end{figure*}

\noindent \textbf{Heterogeneous Data Benchmarks.}
\bchange{Recent benchmarks have begun to evaluate data agents over more diverse data sources. For data analysis scenarios, DABStep~\cite{DBLP:journals/corr/dabstep} provides 450 multi-step analytical tasks over tabular files with domain documentation, \textcolor{black}{KramaBench~\cite{DBLP:journals/corr/abs-2506-06541} targets end-to-end data-to-insight pipelines over data lakes, emphasizing pipeline design and executable workflow construction, and DAComp~\cite{lei2025dacomp} benchmarks enterprise data intelligence workflows with repository-level data engineering pipelines and open-ended analytical tasks.} For data science scenarios, DSBench~\cite{DBLP:conf/iclr/DSBench} provides 540 tasks combining data analysis and modeling, DA-Code~\cite{huang2024dacode} offers 500 code generation tasks with executable environments, BLADE~\cite{gu2024blade} provides 188 decision questions across 12 datasets with notebook interaction, and DS-1000~\cite{Lai2022DS1000} contains 1,000 code generation problems. However, these benchmarks do not comprehensively evaluate the reasoning ability of data agents over heterogeneous data and focus on limited data modalities; current evaluation also relies heavily on LLM-as-judge, which may introduce single-model bias. Regarding benchmark construction, existing approaches either rely on manual curation or domain-specific data~\cite{DBLP:journals/corr/dabstep, gu2024blade}, which does not scale to multi-step agent workflows with iterative tool invocation, or derive tasks from pre-existing code with filtering and cleaning~\cite{DBLP:conf/iclr/DSBench, lei2025dacomp}, which lacks ground-truth guarantees for complex reasoning chains.}

For completeness, we also discuss general agent benchmarks (GAIA, AgentBoard, WebArena, etc.) in Appendix~\ref{appendix:other_benchmarks}.

\section{BENCHMARK OVERVIEW}

\subsection{Target Systems}\label{sec:target_test_systems}

\bchange{FDABench targets data agent systems that process heterogeneous analytical workloads over structured databases $\mathcal{D} = \{D_1, \ldots, D_n\}$ and unstructured sources $\mathcal{K} = \{K_1, \ldots, K_p\}$ encompassing documents and multimedia. Data agents orchestrate multiple tools $\mathcal{T} = \{T_1, \ldots, T_m\}$ through an iterative interaction loop. Given a natural language query $Q$, the agent execution proceeds as follows:}
\begin{align}
s_t &= (Q, \mathcal{D}, \mathcal{K}, h_t), \quad a_t = \pi(s_t), \quad a_t \in \mathcal{T} \cup \{\texttt{final}\} \\
o_t &= \texttt{Exec}(a_t, \mathcal{D}, \mathcal{K}), \quad h_{t+1} = h_t \cup \{(a_t, o_t)\}, \quad h_0 = \emptyset
\end{align}
\bchange{where $s_t$ is the state at step $t$, $h_t$ is the interaction history, $\pi$ is the agent policy (implemented via LLM), $a_t$ is the action (tool invocation or final answer), and $o_t$ is the observation from tool execution. The loop terminates when $a_t = \texttt{final}$ or $t$ reaches budget $B$. FDABench records the action sequence $\{a_0, \ldots, a_T\}$ to compute metrics including Tool F1, Success Rate, latency, and cost.}

\bchange{FDABench evaluates four representative data agent workflow patterns~\cite{LLMDATASurvey, tangllmasanalyst}: Planning, Tool-use, Reflection, and Multi-agent. 
}

\subsection{Design Goals}
\label{sec:designgoal}

\textcolor{black}{As introduced in Sec.~\ref{sec:intro}, evaluating data agents requires addressing four critical aspects: data heterogeneity, task complexity, output diversity, and workflow variability. We operationalize these aspects through Jim Gray's classical benchmark design principles~\cite{DBLP:books/mk/Gray91}, establishing a systematic framework for constructing FDABench.}

\textcolor{black}{\noindent \textbf{Relevance.} FDABench constructs 2,007 heterogeneous analytical tasks by systematically unifying structured databases with diverse unstructured sources (1,600+ PDFs, images, audio, video, and web content), spanning from simple queries to complex multi-step data agent tasks.}

\textcolor{black}{\noindent \textbf{Simplicity.} FDABench standardizes agent responses into three task types: single-choice, multiple-choice, and report generation, transforming variable outputs into structured evaluation formats.}

\textcolor{black}{\noindent \textbf{Portability.} FDABench provides modular interfaces supporting four agent patterns and adapts to different foundation models and tool implementations with minimal configuration, enabling fair comparison across diverse systems~\cite{palimpzestCIDR,patel2024semanticoperators,DBLP:journals/corr/dagent}.}

\textcolor{black}{\noindent \textbf{Scalability.} We design \textit{PUDDING}, an agentic dataset construction framework that \textcolor{black}{performs tree-structured LLM exploration with per-branch self-reflection and iterative expert validation} to efficiently produce 2,007 high-quality tasks, supporting comprehensive data modalities and task types.}

\subsection{Benchmarking Workflow}
As illustrated in Figure~\ref{fig:benchmark_overview}, our benchmarking workflow consists of two primary phases: the Construction Phase and the Evaluation Phase, each designed to systematically generate test cases and assess the capabilities of data agents, respectively.

\noindent \textbf{Construction Phase.} \textcolor{black}{This phase generates benchmark tasks via \textit{PUDDING} (Sec.~\ref{sec:dg}). Given 4,127 candidate queries and their associated databases from verified benchmarks (BIRD, Spider, Spider2, DABStep) paired with 1,600+ curated unstructured files, PUDDING performs tree-structured multi-source exploration with per-branch self-reflection and iterative expert validation (accept / revise / reject), yielding 2,007 high-quality tasks (48.6\% acceptance rate).}

\noindent \textbf{Evaluation Phase.} This phase assesses \textit{Target Data Agent Systems} through multi-stage evaluation. Target systems are integrated into our framework and process FDABench tasks across three categories (single-choice, multiple-choice, and report), generating responses in standardized formats that capture both final answers and intermediate reasoning steps. The \textit{Evaluation Suite} employs multiple complementary metrics to evaluate performance across each task category.

\begin{figure*}
  \centering
  \includegraphics[width=0.79\linewidth]{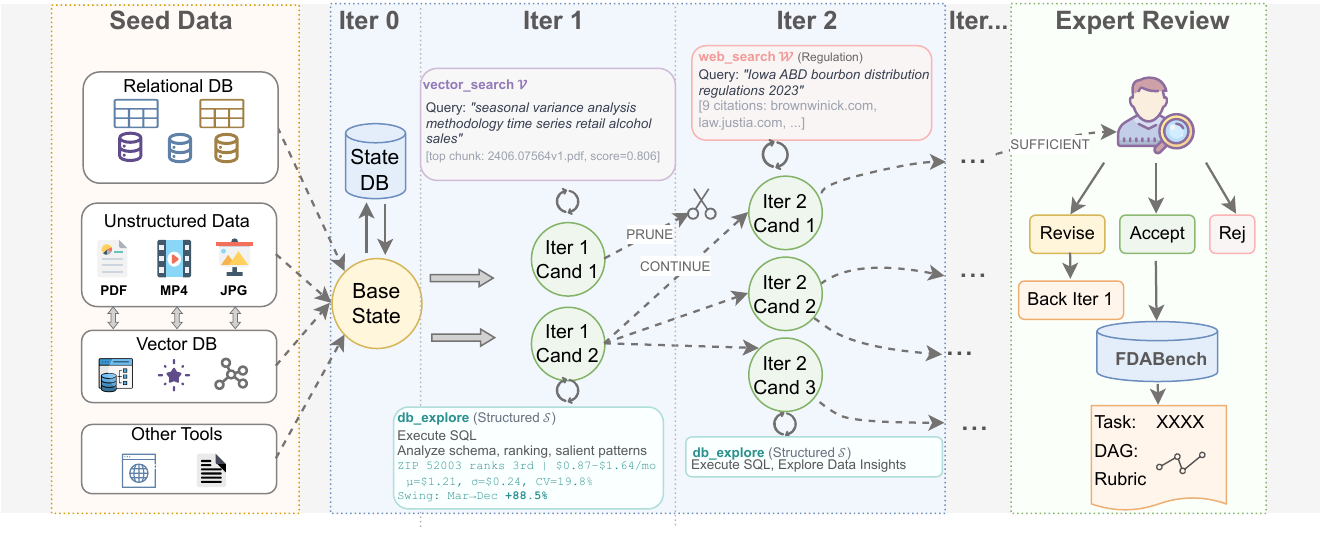}
   \vspace{-0.15in}
\caption{Overview of the \textit{PUDDING} framework.}
   \vspace{-10pt}
  \label{fig:pudding}
\end{figure*}

\section{DATASET CONSTRUCTION}

\bchange{FDABench is constructed via a two-stage pipeline. We first collect heterogeneous data sources (Sec.~\ref{sec:hdc}), and then design \textit{PUDDING} to generate reliable and complex test cases for query analytical tasks on these data sources (Sec.~\ref{sec:dg}).}

\subsection{Heterogeneous Seed Data Collection}
\label{sec:hdc}

\noindent \bchange{\textbf{Structured data.}
We start from four vetted benchmarks (BIRD, Spider, Spider2, and DABStep) containing 23,900+ queries, and manually select 4,127 candidates that reflect realistic analytical agent workloads (e.g., diagnosing anomalies, explaining trends, and attributing causes) rather than one-shot SQL lookups.
Each candidate provides a \textcolor{black}{seed} database instance $D_i \in \mathcal{D}$, its schema, a \textcolor{black}{demonstration} SQL query $s^*$ for database exploration, and executable results from the source benchmark.}

\noindent \bchange{\textbf{Unstructured data.}
To make tasks genuinely multi-source, we curate a complementary unstructured corpus $\mathcal{K} = \{K_1, \ldots, K_p\}$ that provides (i) contextual interpretation (definitions and methodology), (ii) external validation (standards, regulations, market/clinical reports), and (iii) causal explanations that are not derivable from database facts alone.
We collect 1,600+ authoritative files, including academic literature, industry documentation, technical specifications, multimedia content, and enterprise case studies.
Documents are preprocessed into semantically coherent chunks with dense embeddings and indexed via FAISS with provenance metadata, enabling consistent retrieval during construction and frozen-evidence replay during evaluation.}

\subsection{Test Case Construction~\label{sec:dg}}

\bchange{Given a seed tuple $(D_i, q, s^*)$, our goal is to construct a benchmark task that is (i) \emph{verifiable} (execution-grounded ground truth), (ii) \emph{non-trivial} (requires synthesizing $\mathcal{D}$ and $\mathcal{K}$, and optionally web evidence), and (iii) \emph{traceable} (exposes the intended reasoning workflow for tool-sequence/DAG evaluation). We operationalize this goal with two task families and synthesize tasks via \textit{PUDDING}.}

\subsubsection{Task Design.}
\bchange{To comprehensively evaluate data agents over heterogeneous sources, we design two task families that jointly cover objective correctness and open-ended analytical synthesis.}

\noindent \bchange{\textbf{Choice-based tasks.}}
\bchange{We define two objective modes: single-choice (SC) and multiple-choice (MC). Both enforce standardized answer formats while requiring cross-source reasoning and aggregation.
SC emphasizes precise analytics where the correct option is determined by SQL execution plus contextual interpretation, while MC emphasizes synthesis where selecting all correct options requires integrating evidence from both $\mathcal{D}$ and $\mathcal{K}$.
For example, identifying product features that influence customer satisfaction requires combining SQL aggregations on review ratings with unstructured evidence (e.g., review text and product descriptions).}

\noindent \bchange{\textbf{Report-based tasks.}}
\bchange{Report tasks assess analytical writing and reasoning by requiring comprehensive reports grounded in both database evidence and external context. They demand query decomposition, user-intent recognition, and multi-source synthesis.
}

\subsubsection{The Framework of PUDDING}
\label{sec:phase123}

\bchange{We design \textit{PUDDING} as an \emph{agent stateful orchestration graph} \textcolor{black}{$\mathcal{O} = (\mathcal{V}, \mathcal{A}, \sigma, \delta)$}, where \textcolor{black}{$\mathcal{V}$} is a set of functional nodes (deterministic or LLM-powered); \textcolor{black}{$\mathcal{A}$} is the set of conditional transitions governed by runtime predicates; \textcolor{black}{$\sigma$} maps each node $v$ to a per-node state $\sigma_v$ over query context, retrieval artifacts, task drafts, and expert decisions; and $\delta$ defines the transition function.
As shown in Figure~\ref{fig:pudding}, given a seed tuple $(D_i, q, s^*)$, the agent initializes a \emph{Base State} $\sigma_0$ from the heterogeneous data sources (Iter~0) and enters a \textcolor{black}{tree-structured candidate exploration}.
Each iteration $t$ spawns multiple candidate paths that invoke heterogeneous tools (database exploration, vector retrieval, web search); \textcolor{black}{per-branch self-reflection independently evaluates each new leaf $v'$ and issues one of three decisions: \textsc{Sufficient}, \textsc{Continue}, or \textsc{Prune}}.
The loop proceeds autonomously \textcolor{black}{until all frontier leaves are terminal (\textsc{Sufficient} or \textsc{Pruned}) or $N_{max}$ is reached}. \textcolor{black}{All \textsc{Sufficient} terminal paths independently produce task drafts forwarded to expert review}. A stateful design checkpoints all transitions for deterministic replay. The framework executes in three phases.}

\noindent \bchange{\textbf{Phase 1: Context grounding.}}
\bchange{In Iter~0, the agent constructs the Base State $\sigma_0$ by extracting schema $\mathcal{S}$, executing \textcolor{black}{demonstration} SQL $s^*$ on $D_i$ to obtain structured results $\mathcal{R}$, and retrieving demonstration cases $\mathcal{E}$, curated data agent examples stored in a demonstration knowledge base, that guide realistic task synthesis.
In each subsequent iteration $t \!\geq\! 1$, \textcolor{black}{every active frontier node $v$ spawns $b_v$ candidate branches} (Figure~\ref{fig:pudding}), each invoking a tool from web search $\mathcal{W}$~\cite{Perplexity}, vector retrieval, database exploration, or file system search $\mathcal{F}$, conditioned on the current state and patterns discovered in prior iterations.
\textcolor{black}{Per-branch self-reflection labels each new leaf as \textsc{Prune}, \textsc{Continue}, or \textsc{Sufficient}}; retrieval is thus adaptive in both depth and breadth.
For instance, on \textsc{FDA0182} (Iowa liquor sales, Figure~\ref{fig:pudding}), Iter~1 spawns two candidates: \texttt{db\_explore} executes SQL to surface per-capita expenditure patterns but is labeled \textcolor{black}{\textsc{Prune}} (statistical patterns alone lack causal grounding), while \texttt{vector\_search} retrieves a seasonal variance methodology paper and receives \textcolor{black}{\textsc{Continue}}. In Iter~2, the surviving branch further spawns candidates including a \texttt{web\_search} that retrieves Iowa ABD bourbon distribution regulations and reaches \textcolor{black}{\textsc{Sufficient}}, forwarding the task draft to expert review.
All retrieved artifacts $\textcolor{black}{\mathcal{I}_v}$ are deduplicated and frozen with provenance metadata, and each tool call is logged for Task DAG and rubric construction (Algorithm~\ref{alg:hitl}).
The loop terminates when all frontier leaves are terminal or the budget $N_{max}$ is exhausted. \textcolor{black}{Each \textsc{Sufficient} terminal path $\pi_i$ independently produces a task draft for expert review}.}

\noindent \bchange{\textbf{Phase 2: Agent-Expert collaboration.}}
\bchange{The agent synthesizes \textcolor{black}{each \textsc{Sufficient} terminal path's accumulated context $\sigma_{v_i}$} into a task draft \textcolor{black}{$\mathcal{P}_0^{(i)}$}, grounded in $D_i$ and guided by demonstration cases $\mathcal{E}$.
Continuing the example, the agent drafts a report task combining per-capita Bourbon analysis with causal assessment of Iowa's regulatory model on seasonal consumption, enforcing cross-source reasoning over SQL results and external context.
The draft enters an \emph{expert review} gate where six domain experts, following BIRD~\cite{bird} standards, validate that tasks necessitate joint reasoning over $\mathcal{D}$ and $\mathcal{K}$, verify gold answers through independent SQL execution, and improve analytical coherence.
Each expert issues an \textsc{ACCEPT}, \textsc{REVISE}, or \textsc{REJECT} decision. \textsc{REVISE} routes the task back to an earlier iteration with full state context preserved (Figure~\ref{fig:pudding}), enabling targeted refinement without discarding prior evidence. This loop continues until acceptance or iteration exhaustion ($k \geq N_{max}$).}

\noindent \bchange{\textbf{Phase 3: Validation and annotation.}}
\bchange{Tasks that receive \textsc{ACCEPT} in Phase~2 enter a final validation stage. We first apply single-source sufficiency testing: any task solvable by $\mathcal{D}$ alone or $\mathcal{K}$ alone is rejected, ensuring that retained tasks genuinely require cross-source reasoning. Difficulty labels are then assigned following BIRD~\cite{bird} criteria on SQL complexity, source diversity, reasoning depth, and domain knowledge (see full details in the technical report~\cite{FDAbench}), yielding 19.73\% Easy, 33.18\% Medium, 47.09\% Hard (Krippendorff's $\alpha > 0.78$). Finally, the tool calls logged during Phase~1's iterative exploration are compiled into process annotations: each call becomes a typed node and data dependencies form edges, producing a gold Task DAG $\mathcal{G}$ (Appendix~\ref{appendix:dag}); in parallel, we instantiate a weighted evaluation rubric $\mathcal{B}$ with DAG-based chain validation.}

\bchange{\noindent \textbf{Overall quality control.}
\textit{PUDDING} accepts 2,007 tasks from 4,127 candidates (48.6\% acceptance rate). Single-source solvability accounts for nearly half of rejections. These quality controls help keep FDABench tasks verifiable and non-trivial at scale.}
Database, document, and web evidence dominate the workload; images, audio, and video are long-tail sources. 93\% of report tasks span three or more modalities and 63\% exhibit mutual cross-source dependency, where one source's output parameterizes a query to another. Per-branch self-reflection drives modality diversity and reasoning depth, while the single-source sufficiency test drives cross-source dependency (Section~\ref{sec:pudding_ablation}).

\subsection{Evaluation Framework}
\bchange{We design a multi-granularity evaluation framework. For \emph{choice-based tasks}, exact match (EX) measures discrete decision accuracy. For \emph{report-based tasks}, we provide two process-level artifacts:
(i) \emph{Evaluation rubric} $\mathcal{B}$: inspired by rubric-based evaluation practices~\cite{DBLP:journals/corr/abs-2511-07685}, we design a hierarchical scoring template assessing SQL accuracy, external knowledge integration, logical reasoning, and output completeness. The set of active dimensions and chain-validation requirements are adapted by task complexity.
(ii) \emph{Task DAG} $\mathcal{G}$: to evaluate tool equivalence, a problem also observed in~\cite{liu2026toolgate}, we design a directed acyclic graph encoding tool invocations and data dependencies for multi-step tasks, where semantically equivalent tools or alternative execution orders are treated as valid (Appendix~\ref{appendix:dag}).}

\section{EXPERIMENTS}
\label{sec:experiments}
\textcolor{black}{In this section, we first evaluate existing systems on FDABench, then conduct experiments examining each data agent workflow pattern's impact across different dataset difficulty levels and LLMs, and finally provide data agent cost efficiency analysis.}

\subsection{Experimental Setting}

\noindent \textbf{Dataset.} Table~\ref{tab:dataset_stats} summarizes FDABench's task distribution across 139 databases with balanced task categories and a difficulty distribution skewed toward hard tasks.

\noindent \textbf{Environment.} We implement all experiments in Python and conduct experiments on an Ubuntu Server equipped with dual AMD EPYC 9555 64-core processors and 2.2\,TB DDR4 RAM. We utilize OpenRouter~\cite{openrouter} to provide model inference services for our benchmark evaluation. We use DeepSeek-V3.2~\cite{deepseekv32} as the default inference model.

\begin{table}[!t]
\centering
\caption{Dataset Statistics (Total: 2,007 tasks)}
\label{tab:dataset_stats}
\vspace{-10pt}
\footnotesize
\setlength{\tabcolsep}{5pt}
\begin{tabular}{lr@{\hspace{12pt}}lr}
\toprule
\multicolumn{2}{c}{\textbf{By Difficulty}} & \multicolumn{2}{c}{\textbf{By Task Category}} \\
\midrule
Easy   & 396 (19.73\%) & Report          & 668 (33.28\%) \\
Medium & 666 (33.18\%) & Single-Choice   & 579 (28.85\%) \\
Hard   & 945 (47.09\%) & Multiple-Choice & 760 (37.87\%) \\
\bottomrule
\end{tabular}
\vspace{-10pt}
\end{table}

\noindent \textbf{Evaluation Metrics.} Following typical data agent benchmarks~\cite{DBLP:conf/iclr/spider2, DBLP:conf/emnlp/spider1,bird, DBLP:conf/nips/agentboard}, we employ comprehensive evaluation metrics to assess both effectiveness and cost efficiency of each data agent method. We assess effectiveness along three layers. For choice correctness, we utilize Exact Match (EX) on single-choice (EX\_SC) and multiple-choice (EX\_MC) tasks; a single EX score denotes the combined accuracy across both types. For rubric-based report scoring, \bchange{we utilize Rubric Score (RS), a multi-dimensional weighted score assessing SQL accuracy, external knowledge integration, logical reasoning, and completeness} (full rubric in the technical report~\cite{FDAbench}). For DAG-based reasoning trace and tool evaluation, \bchange{we introduce Tool Orchestration Score (TOS), a DAG-based composite metric that evaluates tool invocation correctness with multi-path equivalence, and Tool F1 (TF) for measuring tool invocation precision and recall}, complemented by Success Rate (SR), the ratio of successfully executed tool calls to total tools; DAG conventions are detailed in Appendix~\ref{appendix:dag}. For cost-efficiency evaluation, we utilize data agent end-to-end inference latency (Lat.), external model call (Ext.Call), token cost (Cost), monetary cost, and reasoning tokens to measure data agents' performance across dimensions. We give the formal definitions of our two new metrics, RS and TOS, below.

\noindent\textbf{Rubric Score (RS).} For report task $i$, let $\mathcal{D}_i$ be the active dimensions, drawn from \texttt{SQL\_ACC}, \texttt{EXT\_INTEG}, \texttt{LOG\_REASON}, and \texttt{COMPLETE}, with weights $w_d = 0.25$ and scores $s_d \in [0, 1]$:
\[
\mathrm{RS}_i = \frac{\sum_{d \in \mathcal{D}_i} w_d \cdot s_d}{\sum_{d \in \mathcal{D}_i} w_d}.
\]
\textbf{SQL\_ACC} ($\{0,1\}$) uses exact string match with an LLM fallback for semantic equivalence; \textbf{EXT\_INTEG}, \textbf{LOG\_REASON}, and \textbf{COMPLETE} ($\{0.0, 0.5, 1.0\}$) are LLM-judged ordinal scores for cross-source synthesis, reasoning-chain coherence, and query coverage. Benchmark RS is the mean over all report instances.

\noindent\textbf{Tool Orchestration Score (TOS).} For task $i$ with gold DAG $\mathcal{G}_i = (V_i, E_i)$, TOS is the average of graph coverage and tool-use quality. Graph coverage averages two recalls over the gold DAG: the fraction of required tool nodes ($V^{\text{req}}_i$) and the fraction of critical-path tool nodes ($V^{\text{crit}}_i$) that the agent actually invokes ($V^{\text{match}}_i$), i.e., $\mathrm{GC}_i = \frac{1}{2}\!\left(\frac{|V^{\text{match}}_i \cap V^{\text{req}}_i|}{|V^{\text{req}}_i|} + \frac{|V^{\text{match}}_i \cap V^{\text{crit}}_i|}{|V^{\text{crit}}_i|}\right)$. Tool-use quality $\mathrm{TQ}_i = \mathrm{TF}_i \cdot \mathrm{SS}_i$ multiplies the F1 between gold and agent tool calls ($\mathrm{TF}_i$, with \texttt{ALT\_GROUP} edges crediting equivalent choices) by a soundness indicator $\mathrm{SS}_i \in \{0, 1\}$ that drops to $0$ on any \texttt{HARD\_DEP} ordering violation. The composite is $\mathrm{TOS}_i = \operatorname{avg}(\mathrm{GC}_i,\;\mathrm{TQ}_i)$. Other metrics (EX\_SC, EX\_MC, SR, Lat., Ext.Call, Cost) follow standard Text2SQL and agent benchmark conventions~\cite{DBLP:conf/emnlp/spider1, bird, DBLP:conf/nips/agentboard}.

\noindent \textbf{Evaluated System Implementation.} \textcolor{black}{Since most data agent systems for general analytical queries ( Taiji~\cite{DBLP:journals/corr/taiji}, AOP~\cite{wang2025aop}, AgenticData~\cite{sun2025agenticdata} ) are currently not open-source, we implement them based on their published papers. These reimplementations follow each system's published design and may not fully reproduce its original performance. In addition, the semantic operator query systems and RAG systems lack complete agent decision capabilities, so we use the Planning Agent workflow to equip them with decision capabilities. However, we still find that existing works do not encompass all typical data agent workflows proposed in Section~\ref{sec:target_test_systems}. Therefore, we implement and open-source four typical data agent workflow patterns as additional baselines (see the technical report~\cite{FDAbench}).}

\subsection{Evaluation on Target Systems}\label{sec:effectiveness}

To demonstrate FDABench's generalizability and adaptability, we evaluate three types of data agent systems: (1) general analytical query systems, (2) data agent systems optimized for semantic operator queries, and (3) data agent systems with RAG frameworks.

\subsubsection{General Analytical Query Systems}\label{sec:general}
We evaluate six data agent systems: Taiji~\cite{DBLP:journals/corr/taiji}, AOP~\cite{wang2025aop}, AgenticData~\cite{sun2025agenticdata}, MLE-STAR~\cite{mlestar}, and Teable~\cite{Teable} all use DeepSeek-V3.2~\cite{deepseekv32} as the foundation model, while DeepAnalyze~\cite{deepanalyze} uses its own fine-tuned agentic model with built-in tool orchestration (details in Appendix~\ref{appendix:baselines}).

\noindent \textbf{Performance Analysis.}
\textcolor{black}{As Table~\ref{tab:dataagent_frameworks_performance} shows, architectural distinctions reveal fundamental trade-offs. Tool-use workflows like DeepAnalyze achieve competitive EX scores through fine-tuned agentic models. AOP, a planning workflow with reflection-style operator optimization, outperforms simpler planning workflows across both RS and EX metrics, but at higher computational cost. Multi-agent workflows show varied effectiveness: MLE-STAR achieves moderate performance with good cost efficiency, while AgenticData and Taiji demonstrate competitive results at higher overhead.}

\textcolor{black}{Examining robustness across difficulty levels reveals non-trivial architecture-difficulty interactions. DeepAnalyze suffers the steepest EX\_SC decline (0.64$\to$0.33, $-$48\% from Easy to Hard), indicating that tool-use workflows are sensitive to task complexity despite maintaining stable latency. In contrast, AOP's reflection mechanism yields the most robust tool success rate (SR drops only 10\%, from 0.79 to 0.71), demonstrating that iterative self-correction effectively recovers from tool execution failures. Notably, Teable's RS \textit{increases} with difficulty (0.37$\to$0.47), suggesting that offline planning may better structure analytical reasoning on complex tasks where unconstrained exploration is counterproductive.}

\textcolor{black}{Regarding efficiency, Teable's offline planning achieves the lowest latency and cost with minimal external calls, but shows adaptation limitations as complexity increases. Multi-agent workflows suffer substantial latency penalties with the highest external call counts (Taiji: 26--31 calls). Notably, the cost gap between the most and least efficient architectures exceeds 2.5$\times$, suggesting that architectural selection is a primary cost lever for data agent deployment.}

\begin{table}[!t]
\centering
\caption{Different Data Agent Frameworks' performance on FDABench across difficulty levels (best performance in \textbf{bold}, second-best \underline{underlined}).}
\footnotesize
\setlength{\tabcolsep}{2pt} 
 \scalebox{0.9}{
\begin{tabular}{l|c|cccccccc}
\hline
Method & Diff. & \textcolor{black}{\textbf{RS$\uparrow$}} & \textcolor{black}{\textbf{TOS$\uparrow$}} & \textbf{EX\_SC} & \textbf{EX\_MC} & \textbf{SR} & \textbf{Cost$\downarrow$} & \textbf{Ext.Call$\downarrow$} & \textbf{Lat.$\downarrow$}\\
\hline
\multirow{3}{*}{MLE-STAR} & E & \textcolor{black}{0.42} & \textcolor{black}{0.28} & 0.47 & 0.29 & 0.75 & \underline{10580} & 8.1 & 148 \\
 & M & \textcolor{black}{0.40} & \textcolor{black}{0.25} & 0.38 & 0.24 & 0.67 & \underline{10822} & 8.6 & 149 \\
 & H & \textcolor{black}{0.36} & \textcolor{black}{0.22} & 0.31 & 0.26 & 0.52 & \underline{13472} & 10.5 & 170 \\
\hline
\multirow{3}{*}{Teable} & E & \textcolor{black}{0.37} & \textcolor{black}{0.26} & 0.36 & 0.33 & 0.66 & \textbf{6877} & \textbf{3.5} & \textbf{98} \\
 & M & \textcolor{black}{0.46} & \textcolor{black}{0.25} & 0.35 & 0.28 & 0.66 & \textbf{8658} & \textbf{4.9} & \textbf{131} \\
 & H & \textbf{\textcolor{black}{0.47}} & \textcolor{black}{0.25} & 0.26 & 0.20 & 0.63 & \textbf{8656} & \textbf{5.1} & \textbf{129} \\
\hline
\multirow{3}{*}{DeepAnalyze} & E & \textbf{\textcolor{black}{0.58}} & \textbf{\textcolor{black}{0.34}} & \underline{0.64} & \textbf{0.42} & 0.42 & 18879 & \underline{6.6} & \underline{143} \\
 & M & \textcolor{black}{0.46} & \textbf{\textcolor{black}{0.30}} & \textbf{0.50} & 0.33 & 0.71 & 20836 & \underline{7.0} & \underline{140} \\
 & H & \textcolor{black}{0.38} & \textcolor{black}{0.25} & \underline{0.33} & 0.25 & \underline{0.68} & 24901 & \underline{7.7} & \underline{154} \\
\hline
\multirow{3}{*}{Taiji} & E & \textcolor{black}{0.48} & \textcolor{black}{0.31} & 0.41 & 0.37 & 0.71 & 14013 & 26.4 & 202 \\
 & M & \underline{\textcolor{black}{0.47}} & \underline{\textcolor{black}{0.29}} & 0.32 & 0.27 & 0.67 & 14296 & 28.8 & 221 \\
 & H & \textcolor{black}{0.42} & \underline{\textcolor{black}{0.27}} & 0.27 & 0.25 & 0.65 & 17479 & 31.1 & 255 \\
\hline
\multirow{3}{*}{AOP} & E & \underline{\textcolor{black}{0.56}} & \underline{\textcolor{black}{0.33}} & 0.62 & \underline{0.41} & \textbf{0.79} & 12026 & 15.2 & 165 \\
 & M & \textbf{\textcolor{black}{0.49}} & \textbf{\textcolor{black}{0.30}} & \underline{0.48} & \textbf{0.38} & \textbf{0.76} & 13721 & 17.6 & 189 \\
 & H & \underline{\textcolor{black}{0.46}} & \textbf{\textcolor{black}{0.28}} & \textbf{0.35} & \textbf{0.31} & \textbf{0.71} & 14960 & 19.8 & 225 \\
\hline
\multirow{3}{*}{AgenticData} & E & \textcolor{black}{0.53} & \textcolor{black}{0.30} & \textbf{0.66} & 0.39 & \underline{0.78} & 16007 & 28.5 & 226 \\
 & M & \textcolor{black}{0.44} & \textcolor{black}{0.28} & 0.40 & \underline{0.37} & \underline{0.74} & 16938 & 21.8 & 274 \\
 & H & \textcolor{black}{0.43} & \textcolor{black}{0.26} & 0.31 & \underline{0.28} & \underline{0.68} & 18030 & 25.2 & 298 \\
\hline
\end{tabular}}
\label{tab:dataagent_frameworks_performance}
\vspace{-10pt}
\end{table}

\begin{table}[!t]
\centering
\caption{Data agent with semantic operator performance on FDABench across difficulty levels (best performance in \textbf{bold}, second-best \underline{underlined}).}
\footnotesize
\setlength{\tabcolsep}{2pt}
\begin{tabular}{l|c|cccccccc}
\hline
Method & Diff. & \textcolor{black}{\textbf{RS$\uparrow$}} & \textcolor{black}{\textbf{TOS$\uparrow$}} & \textbf{EX\_SC} & \textbf{EX\_MC} & \textbf{SR} & \textbf{Cost$\downarrow$} & \textbf{Ext.Call$\downarrow$} & \textbf{Lat.$\downarrow$}\\
\hline
\multirow{3}{*}{Lotus} & E & \textbf{\textcolor{black}{0.48}} & \underline{\textcolor{black}{0.30}} & \underline{0.72} & \textbf{0.52} & \textbf{0.89} & \textbf{26447} & \textbf{98.2} & \textbf{642} \\
 & M & \underline{\textcolor{black}{0.50}} & \underline{\textcolor{black}{0.28}} & \underline{0.41} & 0.33 & \textbf{0.66} & \textbf{31387} & \textbf{95.8} & \textbf{658} \\
 & H & \underline{\textcolor{black}{0.52}} & \textbf{\textcolor{black}{0.30}} & 0.34 & 0.14 & \textbf{0.74} & \textbf{33539} & \textbf{97.4} & \textbf{652} \\
\hline
\multirow{3}{*}{Palimpzest} & E & \textcolor{black}{0.46} & \textbf{\textcolor{black}{0.32}} & 0.18 & 0.38 & 0.82 & 41512 & 152.3 & 1015 \\
 & M & \textcolor{black}{0.43} & \textcolor{black}{0.27} & 0.28 & \textbf{0.42} & 0.59 & 47470 & 158.7 & 1039 \\
 & H & \textcolor{black}{0.37} & \textcolor{black}{0.24} & \textbf{0.38} & \textbf{0.18} & 0.68 & 45055 & 154.1 & 1027 \\
\hline
\multirow{3}{*}{DocETL} & E & \underline{\textcolor{black}{0.47}} & \textcolor{black}{0.29} & \textbf{0.79} & \underline{0.51} & \underline{0.87} & \underline{26680} & \underline{101.5} & \underline{669} \\
 & M & \textbf{\textcolor{black}{0.51}} & \textbf{\textcolor{black}{0.30}} & \textbf{0.43} & \underline{0.39} & \underline{0.65} & \underline{32943} & \underline{104.2} & \underline{683} \\
 & H & \textbf{\textcolor{black}{0.53}} & \underline{\textcolor{black}{0.29}} & \underline{0.36} & \underline{0.16} & \underline{0.73} & \underline{35632} & \underline{99.6} & \underline{677} \\
\hline
\end{tabular}
\label{tab:semantic_operator_performance}
\vspace{-10pt}
\end{table}

\begin{table}[!t]
\centering
\caption{Data agent with different RAG methods performance on FDABench across difficulty levels (best performance in \textbf{bold}, second-best \underline{underlined}).}
\footnotesize
\setlength{\tabcolsep}{2pt}
\scalebox{0.9}{
\begin{tabular}{l|c|cccccccc}
\hline
Method & Diff. & \textcolor{black}{\textbf{RS$\uparrow$}} & \textcolor{black}{\textbf{TOS$\uparrow$}} & \textbf{EX\_SC} & \textbf{EX\_MC} & \textbf{SR} & \textbf{Cost$\downarrow$} & \textbf{Ext.Call$\downarrow$} & \textbf{Lat.$\downarrow$}\\
\hline
\multirow{3}{*}{GraphRAG} & E & \textcolor{black}{0.40} & \textcolor{black}{0.26} & 0.45 & 0.28 & 0.65 & 10659 & 23.8 & 291 \\
 & M & \textcolor{black}{0.38} & \textcolor{black}{0.24} & 0.38 & 0.25 & 0.61 & 10760 & 24.1 & 298 \\
 & H & \textcolor{black}{0.36} & \textcolor{black}{0.22} & 0.25 & 0.22 & 0.58 & 11853 & 25.3 & 289 \\
\hline
\multirow{3}{*}{HippoRAG2} & E & \underline{\textcolor{black}{0.42}} & \underline{\textcolor{black}{0.27}} & \underline{0.47} & \underline{0.30} & \underline{0.68} & 12184 & 13.2 & 240 \\
 & M & \underline{\textcolor{black}{0.40}} & \underline{\textcolor{black}{0.25}} & \underline{0.40} & \underline{0.27} & \underline{0.64} & 13108 & 14.1 & 251 \\
 & H & \underline{\textcolor{black}{0.38}} & \underline{\textcolor{black}{0.24}} & \underline{0.28} & \underline{0.24} & \underline{0.61} & 14305 & 13.8 & 248 \\
\hline
\multirow{3}{*}{CARROT} & E & \textbf{\textcolor{black}{0.46}} & \textbf{\textcolor{black}{0.29}} & \textbf{0.52} & \textbf{0.33} & \textbf{0.72} & \underline{3552} & \underline{4.0} & \textbf{197} \\
 & M & \textbf{\textcolor{black}{0.44}} & \textbf{\textcolor{black}{0.27}} & \textbf{0.45} & \textbf{0.30} & \textbf{0.68} & \underline{4062} & \underline{4.3} & \textbf{205} \\
 & H & \textbf{\textcolor{black}{0.41}} & \textbf{\textcolor{black}{0.26}} & \textbf{0.32} & \textbf{0.27} & \textbf{0.65} & \underline{4273} & \underline{5.4} & \textbf{194} \\
\hline
\multirow{3}{*}{NaiveRAG} & E & \textcolor{black}{0.38} & \textcolor{black}{0.24} & 0.41 & 0.25 & 0.62 & \textbf{2806} & \textbf{2.1} & \underline{198} \\
 & M & \textcolor{black}{0.36} & \textcolor{black}{0.23} & 0.35 & 0.22 & 0.58 & \textbf{3307} & \textbf{2.4} & \underline{207} \\
 & H & \textcolor{black}{0.34} & \textcolor{black}{0.21} & 0.24 & 0.20 & 0.55 & \textbf{3486} & \textbf{3.1} & \underline{195} \\
\hline
\end{tabular}
}
\label{tab:rag_methods_performance}
\vspace{-10pt}
\end{table}

\begin{table*}[!t]
    \centering
    \caption{Effectiveness comparison across varying data agent workflow patterns and different LLMs (best performance in \textbf{bold}, second-best \underline{underlined})}
    
    \label{tab:patterns_eff}
    \resizebox{\textwidth}{!}{
    \begingroup
    \setlength{\tabcolsep}{2.5pt}
    \begin{tabular}{l| cccccc | cccccc | cccccc | cccccc}
        \hline
        \multirow{2}{*}{\textbf{Methods}} & \multicolumn{6}{c}{\textbf{Reflection}} & \multicolumn{6}{c}{\textbf{Planning}} & \multicolumn{6}{c}{\textbf{Tool-use}} & \multicolumn{6}{c}{\textbf{Multi-agent}} \\
        \cline{2-25}
        & \textbf{RS$\uparrow$} & \textbf{TOS$\uparrow$} & \textbf{EX$\uparrow$} & \textbf{TF$\uparrow$} & \textbf{SR$\uparrow$} & \textbf{Cost$\downarrow$}
        & \textbf{RS$\uparrow$} & \textbf{TOS$\uparrow$} & \textbf{EX$\uparrow$} & \textbf{TF$\uparrow$} & \textbf{SR$\uparrow$} & \textbf{Cost$\downarrow$}
        & \textbf{RS$\uparrow$} & \textbf{TOS$\uparrow$} & \textbf{EX$\uparrow$} & \textbf{TF$\uparrow$} & \textbf{SR$\uparrow$} & \textbf{Cost$\downarrow$}
        & \textbf{RS$\uparrow$} & \textbf{TOS$\uparrow$} & \textbf{EX$\uparrow$} & \textbf{TF$\uparrow$} & \textbf{SR$\uparrow$} & \textbf{Cost$\downarrow$} \\
        \hline
        GPT-5 & \textcolor{black}{0.418} & \underline{\textcolor{black}{0.331}} & \textbf{0.628} & \textcolor{black}{0.513} & 0.358 & 12331 & \textcolor{black}{0.409} & \textbf{\textcolor{black}{0.412}} & \textbf{0.610} & \textcolor{black}{0.742} & 0.432 & 4430 & \textcolor{black}{0.450} & \textbf{\textcolor{black}{0.392}} & \underline{0.536} & \textcolor{black}{0.606} & 0.386 & \textbf{2587} & \textcolor{black}{0.421} & \underline{\textcolor{black}{0.344}} & \underline{0.622} & \textcolor{black}{0.553} & 0.485 & 11952 \\
        GPT-5-Mini & \textcolor{black}{0.429} & \textcolor{black}{0.321} & \underline{0.553} & \textcolor{black}{0.529} & 0.360 & 12230 & \textcolor{black}{0.412} & \textcolor{black}{0.339} & 0.562 & \textcolor{black}{0.778} & 0.561 & 6242 & \textcolor{black}{0.465} & \underline{\textcolor{black}{0.376}} & \textbf{0.583} & \textcolor{black}{0.574} & 0.213 & 3620 & \textcolor{black}{0.416} & \textbf{\textcolor{black}{0.360}} & 0.581 & \textbf{\textcolor{black}{0.754}} & 0.519 & 15356 \\
        GPT-OSS-120B & \textcolor{black}{0.398} & \textcolor{black}{0.086} & 0.458 & \textcolor{black}{0.152} & 0.615 & \underline{6098} & \textcolor{black}{0.391} & \textcolor{black}{0.299} & 0.438 & \textcolor{black}{0.809} & 0.551 & 6261 & \textcolor{black}{0.400} & \textcolor{black}{0.099} & 0.493 & \textcolor{black}{0.185} & 0.511 & \underline{2730} & \underline{\textcolor{black}{0.422}} & \textcolor{black}{0.180} & 0.549 & \textcolor{black}{0.593} & 0.499 & 20161 \\
        Llama-4-Maverick & \underline{\textcolor{black}{0.521}} & \textcolor{black}{0.094} & 0.305 & \textcolor{black}{0.161} & 0.355 & 9941 & \textbf{\textcolor{black}{0.492}} & \textcolor{black}{0.224} & 0.267 & \textcolor{black}{0.775} & 0.530 & 4425 & \textcolor{black}{0.507} & \textcolor{black}{0.232} & 0.237 & \underline{\textcolor{black}{0.685}} & 0.499 & 7379 & \textcolor{black}{0.407} & \textcolor{black}{0.143} & 0.236 & \textcolor{black}{0.703} & 0.542 & 10106 \\
        Claude-Sonnet-4 & \textcolor{black}{0.438} & \textcolor{black}{0.298} & 0.153 & \textcolor{black}{0.582} & \underline{0.676} & \textbf{5227} & \textcolor{black}{0.385} & \underline{\textcolor{black}{0.345}} & 0.160 & \textcolor{black}{0.796} & \underline{0.620} & \textbf{4179} & \textcolor{black}{0.442} & \textcolor{black}{0.305} & 0.153 & \textcolor{black}{0.588} & 0.676 & 5194 & \textcolor{black}{0.398} & \textcolor{black}{0.312} & 0.296 & \underline{\textcolor{black}{0.738}} & 0.554 & 9202 \\
        DeepSeek-R1-0528 & \textcolor{black}{0.412} & \textcolor{black}{0.285} & 0.439 & \underline{\textcolor{black}{0.625}} & 0.442 & 26015 & \textcolor{black}{0.378} & \textcolor{black}{0.318} & 0.448 & \textcolor{black}{0.758} & \textbf{0.622} & 14341 & \textcolor{black}{0.415} & \textcolor{black}{0.268} & 0.373 & \textcolor{black}{0.542} & \textbf{0.808} & 9849 & \textcolor{black}{0.395} & \textcolor{black}{0.248} & 0.511 & \textcolor{black}{0.498} & 0.533 & 14900 \\
        DeepSeek-V3.2 & \textcolor{black}{0.505} & \textcolor{black}{0.325} & 0.295 & \textbf{\textcolor{black}{0.688}} & 0.290 & 8934 & \textcolor{black}{0.294} & \textcolor{black}{0.322} & 0.310 & \textbf{\textcolor{black}{0.883}} & 0.541 & \underline{4224} & \underline{\textcolor{black}{0.525}} & \textcolor{black}{0.299} & 0.225 & \textbf{\textcolor{black}{0.767}} & 0.660 & 5387 & \textcolor{black}{0.395} & \textcolor{black}{0.340} & 0.458 & \textcolor{black}{0.655} & 0.482 & \textbf{6514} \\
        Gemini-2.5-Flash & \textcolor{black}{0.425} & \textcolor{black}{0.268} & 0.261 & \textcolor{black}{0.595} & 0.469 & 26362 & \textcolor{black}{0.408} & \textcolor{black}{0.255} & 0.349 & \textcolor{black}{0.628} & 0.489 & 6250 & \textcolor{black}{0.418} & \textcolor{black}{0.262} & 0.202 & \textcolor{black}{0.578} & 0.462 & 6945 & \textcolor{black}{0.402} & \textcolor{black}{0.258} & 0.385 & \textcolor{black}{0.565} & 0.463 & 24397 \\
        Gemini-2.5-Pro & \textcolor{black}{0.438} & \textcolor{black}{0.215} & 0.267 & \textcolor{black}{0.468} & 0.227 & 33066 & \textcolor{black}{0.385} & \textcolor{black}{0.198} & 0.191 & \textcolor{black}{0.398} & 0.373 & 9274 & \textcolor{black}{0.455} & \textcolor{black}{0.222} & 0.152 & \textcolor{black}{0.435} & 0.357 & 4955 & \textcolor{black}{0.395} & \textcolor{black}{0.208} & 0.369 & \textcolor{black}{0.442} & 0.431 & 19437 \\
        Qwen3-30B-A3B & \textcolor{black}{0.422} & \textcolor{black}{0.272} & 0.224 & \textcolor{black}{0.539} & 0.529 & 30592 & \textcolor{black}{0.442} & \textcolor{black}{0.278} & \underline{0.605} & \textcolor{black}{0.811} & 0.514 & 7976 & \textcolor{black}{0.419} & \textcolor{black}{0.269} & 0.147 & \textcolor{black}{0.558} & \underline{0.681} & 8120 & \textcolor{black}{0.403} & \textcolor{black}{0.278} & \textbf{0.627} & \textcolor{black}{0.596} & \underline{0.588} & 15443 \\
        Kimi-K2.5 & \textcolor{black}{0.437} & \textbf{\textcolor{black}{0.361}} & 0.305 & \textcolor{black}{0.618} & \textbf{0.707} & 15855 & \underline{\textcolor{black}{0.445}} & \textcolor{black}{0.323} & 0.579 & \textcolor{black}{0.868} & 0.448 & 4386 & \textcolor{black}{0.459} & \textcolor{black}{0.332} & 0.288 & \textcolor{black}{0.674} & 0.622 & 5244 & \textcolor{black}{0.409} & \textcolor{black}{0.323} & 0.584 & \textcolor{black}{0.686} & \textbf{0.644} & \underline{7627} \\
        CodeStral-2508 & \textbf{\textcolor{black}{0.606}} & \textcolor{black}{0.210} & 0.222 & \textcolor{black}{0.369} & 0.277 & 13387 & \textcolor{black}{0.416} & \textcolor{black}{0.314} & 0.198 & \underline{\textcolor{black}{0.870}} & 0.491 & 4382 & \textbf{\textcolor{black}{0.670}} & \textcolor{black}{0.235} & 0.261 & \textcolor{black}{0.417} & 0.473 & 3975 & \textbf{\textcolor{black}{0.561}} & \textcolor{black}{0.330} & 0.535 & \textcolor{black}{0.724} & 0.442 & 9160 \\
        \hline
    \end{tabular}
    \endgroup
    }
    \vspace{-10pt}
\end{table*}

\subsubsection{Semantic Operator Query Systems}\label{sec:semantic_exp}
\textcolor{black}{Since semantic operator systems (Palimpzest~\cite{palimpzestCIDR}, LOTUS~\cite{patel2024semanticoperators}, DocETL~\cite{shankar2024docetl}) lack complete agent self-decision capabilities, we employ the Planning Agent workflow with semantic operators replacing data processing components, using DeepSeek-V3.2~\cite{deepseekv32}.}

\begin{table}
\centering
\caption{Data Agent workflows performance across difficulty levels (best performance in \textbf{bold}, second-best \underline{underlined})}
\footnotesize
\setlength{\tabcolsep}{2pt}
\begin{tabular}{l|c|ccccccc}
\hline
Method & Diff. & \textcolor{black}{\textbf{RS$\uparrow$}} & \textcolor{black}{\textbf{TOS$\uparrow$}} & \textbf{EX$\uparrow$} & \textcolor{black}{\textbf{TF$\uparrow$}} & \textbf{SR$\uparrow$} & \textbf{Cost$\downarrow$}\\
\hline
\multirow{3}{*}{Reflection} & E & \textcolor{black}{0.452} & \textcolor{black}{0.272} & 0.361 & \textbf{\textcolor{black}{0.538}} & 0.405 & 8925 \\
 & M & \underline{\textcolor{black}{0.448}} & \textcolor{black}{0.268} & \underline{0.333} & \textbf{\textcolor{black}{0.521}} & 0.403 & 9388 \\
 & H & \underline{\textcolor{black}{0.435}} & \textcolor{black}{0.255} & \underline{0.240} & \textbf{\textcolor{black}{0.485}} & 0.280 & 11168 \\
\hline
\multirow{3}{*}{Planning} & E & \textcolor{black}{0.398} & \underline{\textcolor{black}{0.295}} & \underline{0.457} & \underline{\textcolor{black}{0.512}} & \underline{0.606} & \textbf{3519} \\
 & M & \textcolor{black}{0.382} & \textcolor{black}{0.258} & 0.285 & \underline{\textcolor{black}{0.486}} & \underline{0.515} & \textbf{4698} \\
 & H & \textcolor{black}{0.375} & \textcolor{black}{0.242} & 0.184 & \underline{\textcolor{black}{0.461}} & \underline{0.498} & \textbf{5186} \\
\hline
\multirow{3}{*}{Tool-use} & E & \underline{\textcolor{black}{0.468}} & \textbf{\textcolor{black}{0.298}} & 0.348 & \textcolor{black}{0.425} & \textbf{0.669} & \underline{4099} \\
 & M & \textcolor{black}{0.432} & \textbf{\textcolor{black}{0.275}} & 0.217 & \textcolor{black}{0.398} & \textbf{0.665} & \underline{5472} \\
 & H & \textcolor{black}{0.428} & \underline{\textcolor{black}{0.268}} & 0.191 & \textcolor{black}{0.372} & \textbf{0.656} & \underline{6683} \\
\hline
\multirow{3}{*}{Multi-agent} & E & \textbf{\textcolor{black}{0.485}} & \textbf{\textcolor{black}{0.298}} & \textbf{0.566} & \textcolor{black}{0.478} & 0.498 & 6084 \\
 & M & \textbf{\textcolor{black}{0.465}} & \underline{\textcolor{black}{0.272}} & \textbf{0.442} & \textcolor{black}{0.452} & 0.489 & 6551 \\
 & H & \textbf{\textcolor{black}{0.448}} & \textbf{\textcolor{black}{0.275}} & \textbf{0.265} & \textcolor{black}{0.425} & 0.470 & 6915 \\
\hline
\end{tabular}
\label{tab:workflow_patterns_by_difficulty}
  \vspace{-10pt}
\end{table}

\noindent \textbf{Performance Analysis.}
As Table~\ref{tab:semantic_operator_performance} shows, LOTUS and DocETL demonstrate better performance with consistent external model calls (about 100) and moderate latency (640-680s), while Palimpzest requires higher overhead (150+ calls). LOTUS and DocETL outperform the best general analytical baseline on hard report tasks (e.g., DocETL 0.53 vs.\ best general 0.47), though not on easy tasks where tool-use systems like DeepAnalyze (0.58) lead. These results reveal a fundamental trade-off where semantic operator systems prioritize semantic understanding at 3-7$\times$ computational cost, suggesting future data agent design must balance semantic depth with operational efficiency.

\subsubsection{RAG Framework Systems}\label{sec:rag_exp}

FDABench also supports data agents with RAG frameworks through interface and dataset compatibility. Since RAG frameworks lack agent self-decision capabilities, we equip them with the Planning workflow using the same foundation model (DeepSeek-V3.2~\cite{deepseekv32}) as in Section~\ref{sec:semantic_exp}. We evaluate four RAG frameworks: CARROT~\cite{DBLP:journals/corr/CORAG}, GraphRAG~\cite{GraphRAG}, HippoRAG2~\cite{DBLP:conf/nips/hipporag1, DBLP:journals/corr/hipporag2}, and NaiveRAG~\cite{naiverag}. Detailed system descriptions are provided in Appendix~\ref{appendix:baselines}.

\noindent \textbf{Performance Analysis.}
As Table~\ref{tab:rag_methods_performance} shows, CARROT achieves the best RS across all difficulty levels at only one-third the cost of GraphRAG, demonstrating that cost-aware retrieval optimization outperforms brute-force graph traversal. GraphRAG's exhaustive exploration (over 23 external calls) yields diminishing returns, underperforming the more efficient HippoRAG2. Compared to general analytical systems in Section~\ref{sec:general}, all RAG methods show lower RS and TOS, suggesting that retrieval augmentation cannot substitute for agent-level tool orchestration in heterogeneous data analytics.

\noindent \textbf{Summary.} \textcolor{black}{The above experiments demonstrate trade-offs in data agent architecture selection: complex workflows improve quality at higher cost, while simpler approaches offer efficiency with moderate quality reduction.}

\begin{figure}[!t]
  \centering
  \begin{subfigure}[b]{0.23\textwidth}
    \centering
    \includegraphics[width=\linewidth]{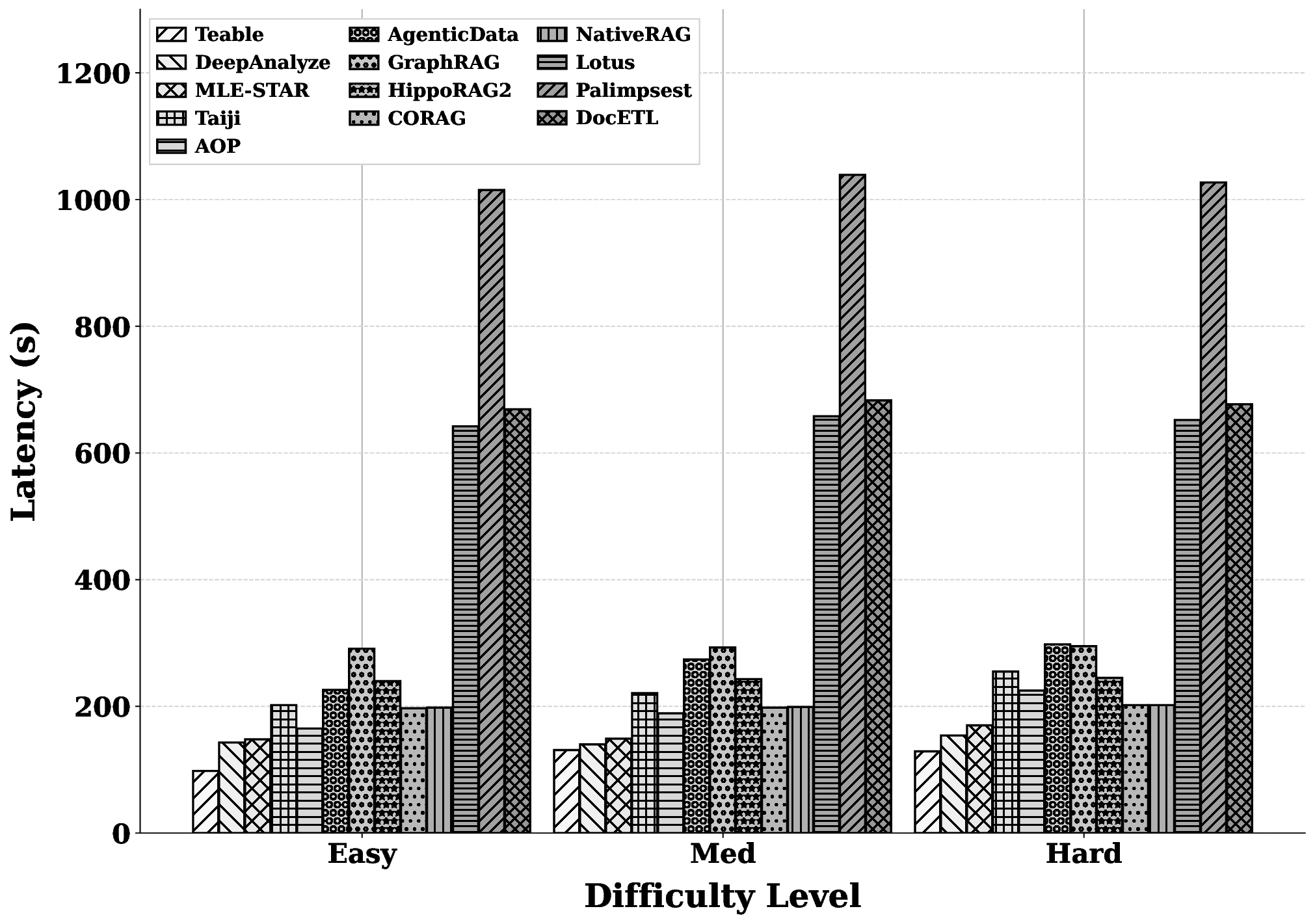}
    \caption{Target Systems Efficiency}
    \label{fig:efficiency_model}
  \end{subfigure}
  \hfill
  \begin{subfigure}[b]{0.23\textwidth}
    \centering
    \includegraphics[width=\linewidth]{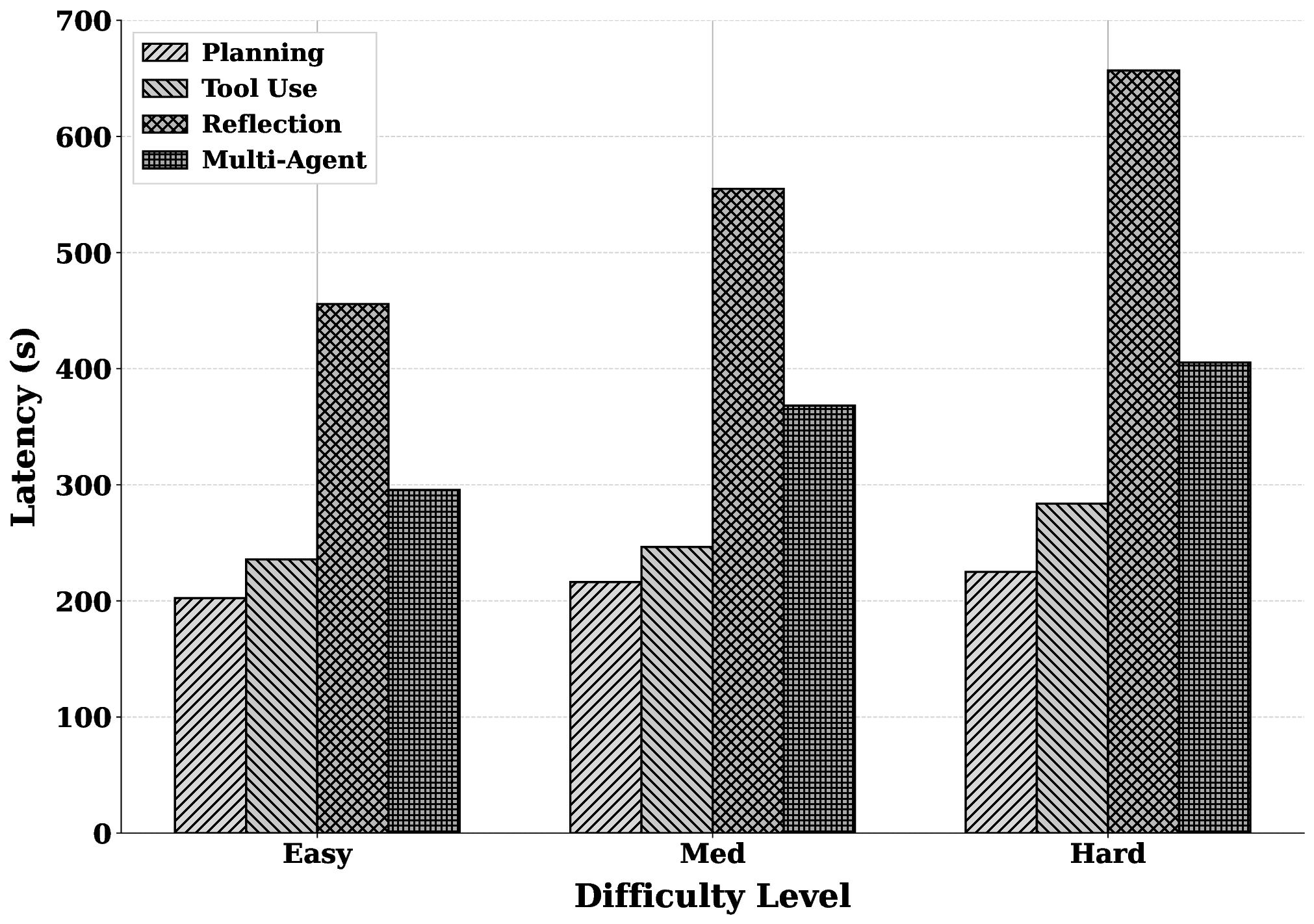}
    \caption{Agent Workflows Efficiency}
    \label{fig:agent_efficiency_comparison}
  \end{subfigure}
  \vspace{-10pt}
  \caption{Computational Efficiency Comparison}
  \label{fig:efficiency_comparison}
\end{figure}

\subsection{Impact of Data Agent Workflow Patterns
}\label{sec:workflow_patterns_impact}
To comprehensively cover all typical workflows, we evaluate 12 foundation LLMs across four workflow patterns and three difficulty levels. The selected models include: 
Claude-Sonnet-4~\cite{claude}, Gemini series~\cite{gemini}, GPT-5 series~\cite{GPT}, GPT-OSS-120B~\cite{gptoss}, Llama-4-Maverick~\cite{llama4}, DeepSeek series~\cite{deepseekv32}, Qwen series~\cite{qwen}, Kimi-K2.5~\cite{kimik2}, and CodeStral~\cite{codestral}.

\noindent \textbf{Performance Across LLMs.}
As shown in Table~\ref{tab:patterns_eff}, frontier reasoning models such as GPT-5 series achieve leading performance across workflows, while code-specialized models such as CodeStral demonstrate superior rubric scores under Reflection and Tool-use workflows. This capability dichotomy suggests that strong reasoning capabilities benefit discrete decision tasks, whereas code-oriented pretraining enhances structured analytical report generation.

Models designed for agentic scenarios such as Kimi-K2.5 exhibit significant performance improvements under Multi-agent and Reflection patterns, with comparatively smaller gains under simpler workflows. This indicates that agentic-oriented design enables more effective utilization of iterative refinement and collaborative mechanisms. Additionally, most LLMs achieve consistently high tool F1 scores under Planning workflows due to explicit plan decomposition, while Tool-use workflows show polarized performance, suggesting that direct tool invocation amplifies inherent differences in LLMs' tool utilization capabilities. We further analyze efficiency characteristics in Section~\ref{sec:costefficiency}.

\begin{figure}[!t]
  \centering
  \includegraphics[width=0.9\linewidth]{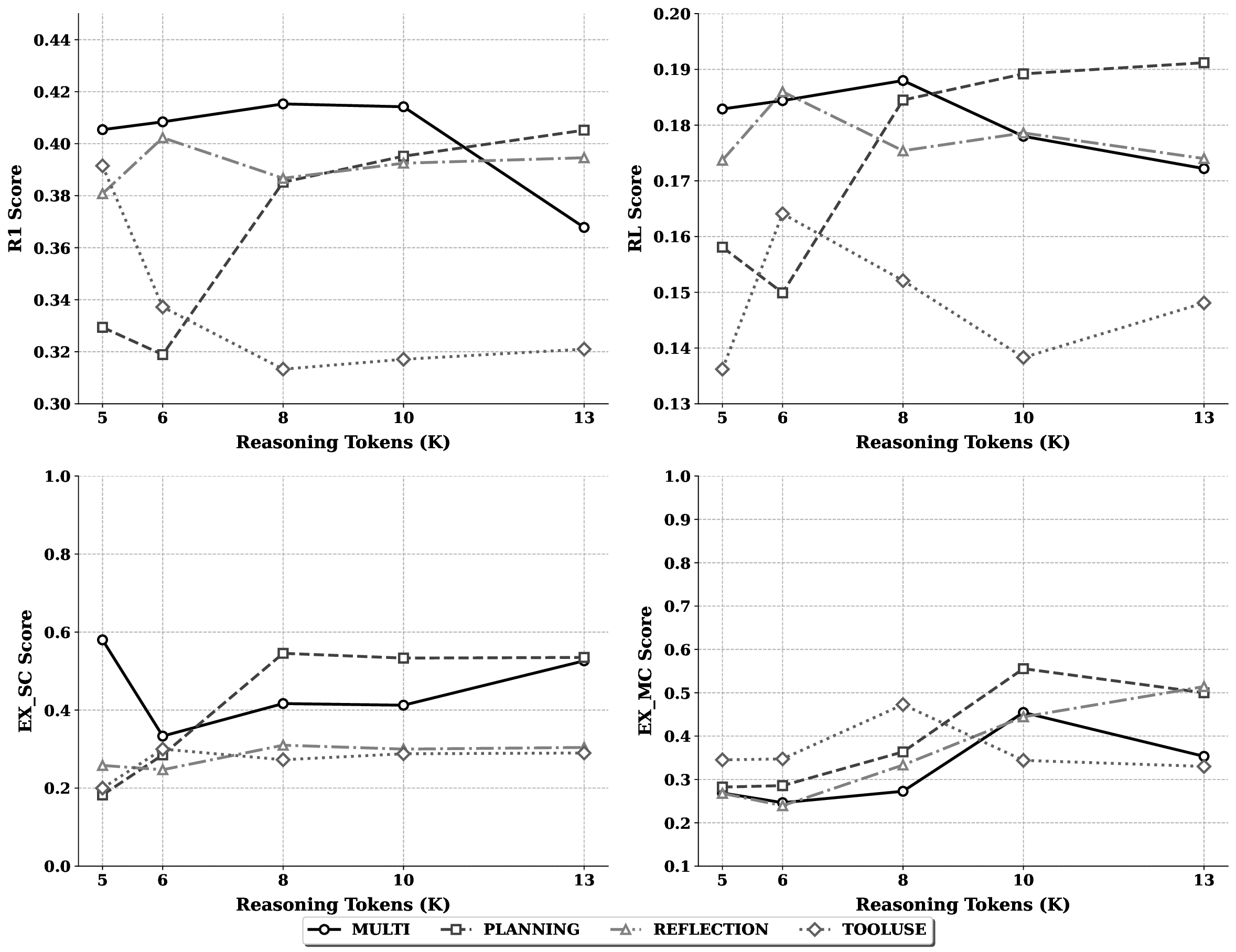}
  \vspace{-10pt}
  \caption{Reasoning Token Scaling Analysis of Different Agent Patterns}
  \vspace{-14pt}
  \label{fig:scaling_law_reasoning_all_curves}
\end{figure}

\noindent \textbf{Robustness to Task Complexity.}
As shown in Table~\ref{tab:workflow_patterns_by_difficulty}, the Planning workflow demonstrates superior cost-efficiency but exhibits significant performance degradation on hard tasks, similar to Teable's behavior in Section~\ref{sec:general}. In contrast, Multi-agent workflows show remarkable robustness. Tool-use workflows maintain high success rates across all difficulties with minimal degradation. Concretely, Reflection's SR falls 31\% (0.405$\to$0.280) from Easy to Hard, whereas Multi-agent's SR holds near $0.5$ across all difficulty levels (0.498$\to$0.470), supplying a per-workflow view of where tool execution itself starts to fail.

\begin{table}[!t]
\centering
\caption{Breakdown latency across different agent patterns on various difficulty levels}
\footnotesize
\resizebox{\linewidth}{!}{
\begin{tabular}{c|c|cccc|c}
\hline
Method & Difficulty & Decision/s & Execute/s & Retry/s & Generate/s & Total/s\\
\hline
\multirow{3}{*}{Planning}
 & Easy & 55.04 (27.19\%) & 74.34 (36.73\%) & 6.02 (2.97\%) & 67.00 (33.10\%) & 202.40 \\
 & Med  & 59.80 (27.65\%) & 77.89 (36.02\%) & 3.29 (1.52\%) & 75.27 (34.81\%) & 216.25 \\
 & Hard & 63.70 (28.31\%) & 85.12 (37.84\%) & 3.81 (1.69\%) & 72.34 (32.16\%) & 224.97 \\
\hline
\multirow{3}{*}{Tool Use}
 & Easy & 86.83 (36.83\%) & 71.55 (30.35\%) & 0.60 (0.25\%) & 76.78 (32.57\%) & 235.76 \\
 & Med  & 98.71 (40.05\%) & 79.91 (32.42\%) & 1.82 (0.74\%) & 66.05 (26.80\%) & 246.49 \\
 & Hard & 119.53 (42.13\%) & 88.39 (31.15\%) & 0.77 (0.27\%) & 75.03 (26.45\%) & 283.72 \\
\hline
\multirow{3}{*}{Reflection}
 & Easy & 125.30 (27.49\%) & 149.03 (32.70\%) & 119.77 (26.28\%) & 61.63 (13.52\%) & 455.73 \\
 & Med  & 183.01 (32.98\%) & 158.65 (28.59\%) & 148.18 (26.70\%) & 65.08 (11.73\%) & 554.92 \\
 & Hard & 233.20 (35.50\%) & 165.00 (25.12\%) & 189.20 (28.81\%) & 69.41 (10.57\%) & 656.81 \\
\hline
\multirow{3}{*}{Multi-agent}
 & Easy & 82.79 (28.00\%) & 111.02 (37.55\%) & 34.75 (11.75\%) & 67.11 (22.70\%) & 295.67 \\
 & Med  & 135.93 (36.92\%) & 123.80 (33.63\%) & 36.60 (9.94\%) & 71.84 (19.51\%) & 368.17 \\
 & Hard & 166.82 (41.16\%) & 119.91 (29.58\%) & 42.44 (10.47\%) & 76.17 (18.79\%) & 405.34 \\
\hline
\end{tabular}}
\label{tab:latency_breakdown_by_pattern}
\vspace{-10pt}
\end{table}

\begin{figure}[!t]
  \centering
  \begin{subfigure}[t]{0.23\textwidth}
    \centering
    \includegraphics[width=\textwidth]{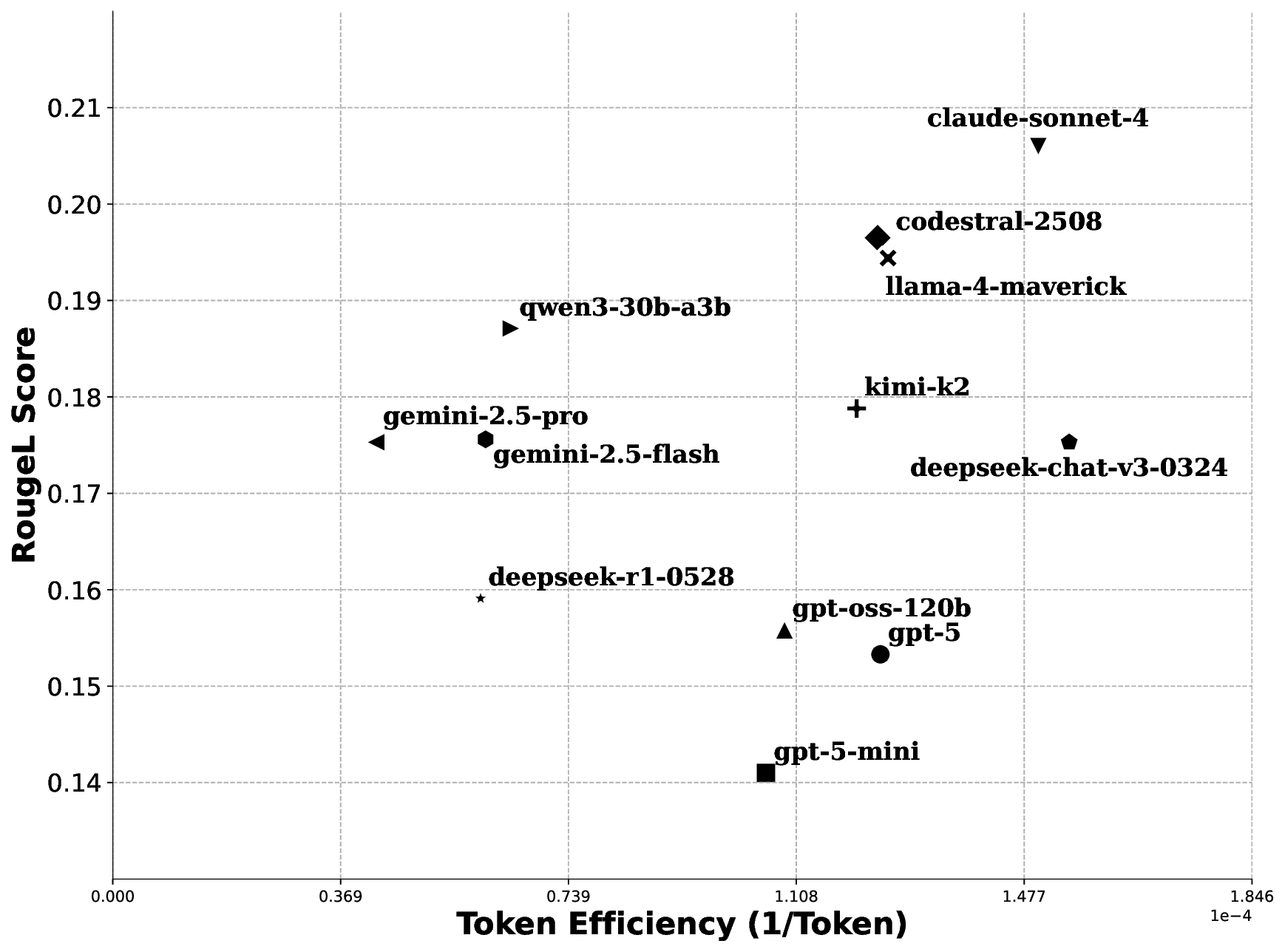}
    \captionsetup{belowskip=2pt,aboveskip=0pt}
    \caption{Token vs. quality}
    \label{fig:token_quality}
  \end{subfigure}
  \hfill
  \begin{subfigure}[t]{0.23\textwidth}
    \centering
    \includegraphics[width=\textwidth]{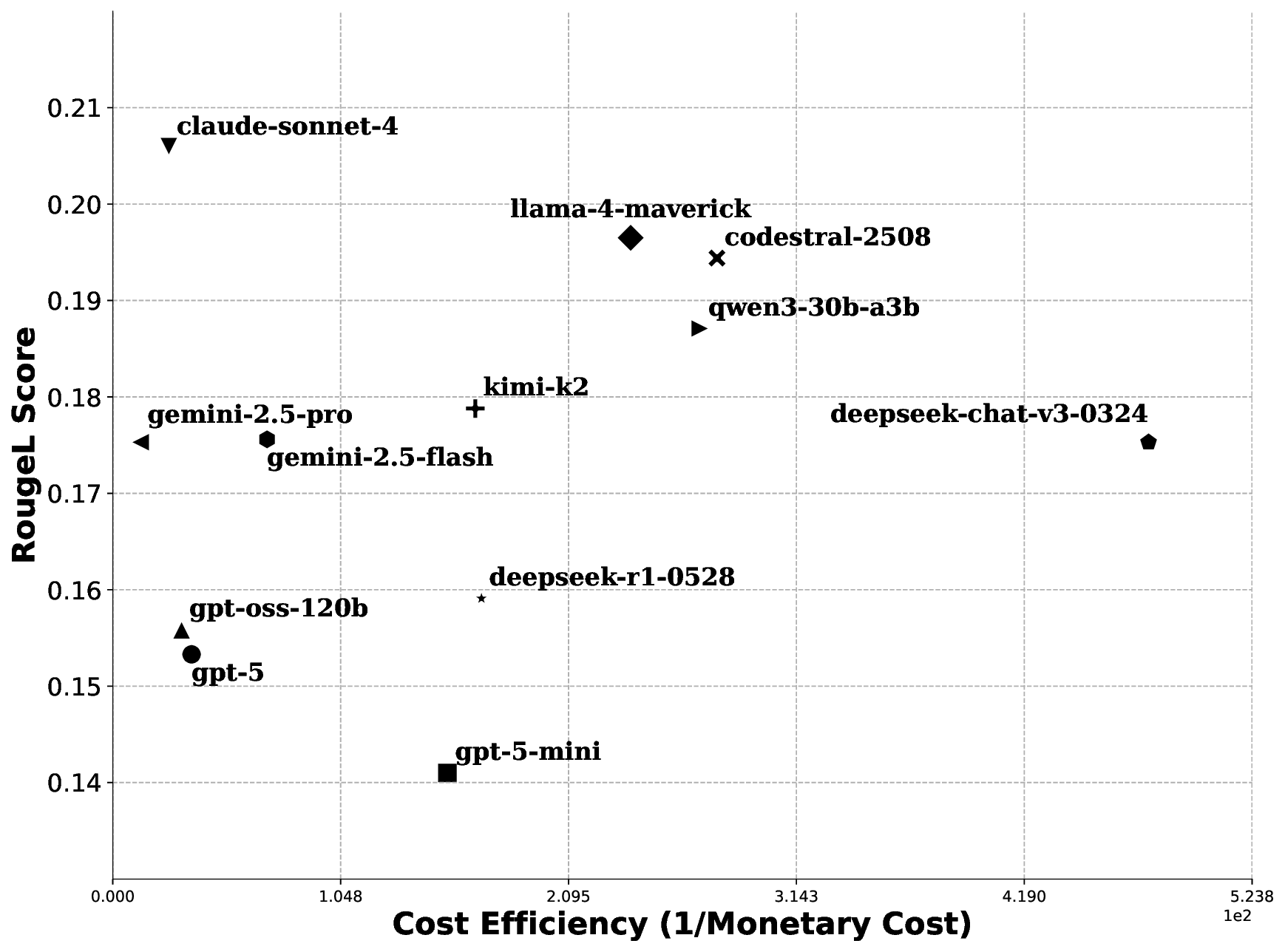}
    \captionsetup{belowskip=2pt,aboveskip=0pt}
    \caption{\small{Monetary cost vs. quality}}
    \label{fig:price_quality}
  \end{subfigure}
  \vspace{-15pt}
  \caption{Performance trade-offs across token, cost, and quality dimensions}
  \label{fig:cost_quality_tradeoff}
\end{figure}

\subsection{Cost Efficiency Analysis}\label{sec:costefficiency}

\noindent \textbf{End-to-end Efficiency.}
We characterize efficiency on the choice-based tasks with a fixed LLM (DeepSeek-V3.2) as shown in Figure~\ref{fig:efficiency_comparison}. Planning workflows maintain the lowest latency across all difficulty levels, while Reflection workflows exhibit 2--3$\times$ higher latency with the steepest growth. On both easy and hard tasks, the slowest general analytical system (AgenticData) is approximately 2.3$\times$ slower than the fastest (Teable), indicating a consistent efficiency gap across difficulties underscoring architecture choice as a primary cost lever.

\textcolor{black}{At the system level, latency scaling patterns further diverge. Teable's latency remains nearly flat across difficulty levels, with hard-task latency actually lower than medium-task, indicating the stability of pre-generated execution plans. DeepAnalyze similarly maintains stable latency despite significant accuracy degradation, suggesting that its domain-specialized agentic model, which is specifically fine-tuned for data science scenarios, preserves execution efficiency even when reasoning quality deteriorates on harder tasks. Notably, latency and token cost do not always scale together: DeepAnalyze's token cost grows 32\% from Easy to Hard while latency stays flat, indicating that harder tasks require denser per-call reasoning rather than additional calls. This suggests that latency-based metrics alone may underestimate the true computational burden of complex tasks.}

\noindent \textbf{Efficiency Breakdown.} We decompose each workflow into Decision, Execute, Retry, and Generate steps (Table~\ref{tab:latency_breakdown_by_pattern}). Retry time is the key differentiator: Reflection allocates 26-29\% to retries, while Planning keeps it below 3\%. As difficulty increases, Reflection's retry proportion grows but Planning's remains stable, indicating that pre-planned decomposition effectively amortizes complexity-induced overhead. Multi-agent workflows show rising decision-time proportion (28\%$\to$41\%), reflecting coordination costs that scale with task complexity.

\noindent \textbf{Reasoning Token Scaling Analysis.}
As shown in Figure~\ref{fig:scaling_law_reasoning_all_curves}, reasoning token scaling benefits are highly workflow-dependent. Tool-use and Reflection workflows demonstrate consistent quality improvements with more reasoning tokens, while Planning and Multi-agent workflows exhibit erratic or flat returns. This asymmetry suggests that reasoning tokens are most effective when the workflow structure provides opportunities for iterative self-correction, and points to workflow-adaptive token budgeting as a promising direction.

\noindent \textbf{Cost-Quality Trade-offs.}
Figure~\ref{fig:cost_quality_tradeoff} reveals that Claude-Sonnet-4 (high quality) and DeepSeek-V3.2 (low cost) form the Pareto frontier. Most other models cluster in a suboptimal region, suggesting that simply scaling model size does not guarantee cost-effective agent performance; architecture-aware model selection is essential.

\subsection{Benchmark Reliability}\label{sec:pudding_ablation}\label{sec:judge_reliability}

\noindent\textbf{Construction reliability.} We conduct an ablation study of PUDDING, disabling one component at a time and rerunning construction on the same 200 sampled seeds (Table~\ref{tab:pudding_ablation}). The components act on different axes: per-branch self-reflection drives modality count and reasoning depth, so removing it costs the most on both; the single-source sufficiency test enforces cross-source dependency, so disabling it sharply reduces the share of two-source tasks while leaving modality breadth intact; expert revision repairs reasoning and realism, lifting acceptance and depth without adding new sources. The Naive variant removes all three and degrades every axis at once, indicating the components are complementary rather than substitutes.

\begin{table}[!t]
\centering
\footnotesize
\caption{PUDDING component ablation on 200 sampled seeds: modalities per task, reasoning depth, \% with cross-source dependency, expert acceptance \%.}
\label{tab:pudding_ablation}
\setlength{\tabcolsep}{4pt}
\begin{tabular}{lcccc}
\toprule
\textbf{Variant} & \textbf{\#Mod.}\,$\uparrow$ & \textbf{Depth}\,$\uparrow$ & \textbf{X-Dep \%}\,$\uparrow$ & \textbf{Acc.\ \%} \\
\midrule
Full PUDDING                  & 3.1 & 3.5 & 63.0 & 48.6 \\
w/o Per-branch Reflection     & 2.3 & 2.5 & 42.0 & 35.5 \\
w/o Expert Revision           & 2.5 & 2.8 & 54.0 & 39.0 \\
w/o Sufficiency Test          & 2.9 & 3.3 & 41.0 & 43.0 \\
Naive (all removed)           & 1.2 & 1.3 & 18.0 & 24.5 \\
\bottomrule
\end{tabular}
\end{table}

\noindent\textbf{Evaluation alignment.} We design our LLM judge with Gemini-3-Flash (\texttt{google/gemini-3-flash-preview}), constrained ordinal scoring on $\{0.0, 0.5, 1.0\}$, citation-grounded prompts, dimensional separation, and exact-match SQL verification with LLM fallback. We conduct a judge validation study where three experts independently score 200 sampled tasks. As shown in Table~\ref{tab:human_llm_agreement}, the judge shows high agreement with expert consensus across task categories, with only a small human--LLM gap on report tasks and preserved system rankings. A per-dimension breakdown is provided in the technical report~\cite{FDAbench}.

\begin{table}[t]
\centering
\caption{Human vs.\ LLM judge agreement across task categories (200 sampled tasks).}
\label{tab:human_llm_agreement}
\footnotesize
\resizebox{\linewidth}{!}{%
\begin{tabular}{llccc}
\toprule
\textbf{Task Category} & \textbf{Comparison} & \textbf{Krippendorff's $\alpha$}\,$\uparrow$ & \textbf{ICC(A,1)}\,$\uparrow$ & \textbf{$\tau_b$}\,$\uparrow$ \\
\midrule
\multirow{2}{*}{Report} & Human vs.\ Human & 0.81 & 0.84 & 0.95 \\
 & Human vs.\ LLM & 0.76 & 0.79 & 0.92 \\
\midrule
\multirow{2}{*}{Single-Choice} & Human vs.\ Human & 0.94 & 0.96 & 0.98 \\
 & Human vs.\ LLM & 0.93 & 0.95 & 0.97 \\
\midrule
\multirow{2}{*}{Multiple-Choice} & Human vs.\ Human & 0.88 & 0.91 & 0.96 \\
 & Human vs.\ LLM & 0.85 & 0.87 & 0.94 \\
\bottomrule
\end{tabular}
}
\end{table}

\section{KEY TAKEAWAYS}
We analyze the experiments above and summarize three design takeaways for data agents. Each connects a recurring failure mode, observed consistently across systems and workflow patterns, to a concrete design guideline for future systems:

\noindent \bchange{\textbf{Takeaway 1: Analytical Pipeline Topology Matters More Than Query Complexity.}}
\bchange{For data agents, task difficulty is not solely determined by SQL complexity or domain knowledge, but by the topological structure of the analytical pipeline. DAG structural features such as deep dependency chains, fork-join branching, and validation-dense subgraphs correlate more closely with agent performance than nominal difficulty tiers (Appendix~\ref{appendix:dag}). As shown in Table~\ref{tab:workflow_patterns_by_difficulty}, Planning achieves higher EX than Reflection on easy tasks (0.457 vs.\ 0.361), yet this ranking \textit{reverses} on hard tasks (0.184 vs.\ 0.240), indicating that static plans collapse under complex topologies while adaptive architectures degrade gracefully.}
Static plans lack recovery branches; DAG, SQL, and domain knowledge co-vary along the difficulty axis.

\noindent \bchange{\textbf{Takeaway 2: Tools Are Easy to Call but Hard to Orchestrate: the Long Tail Decides.}}
\bchange{We observe a clear separation between tool execution reliability and end-to-end task accuracy. As shown in Table~\ref{tab:workflow_patterns_by_difficulty}, Tool-use workflows achieve the highest SR across all difficulty levels (0.669$\to$0.656, merely 2\% degradation), yet trail Reflection and Multi-agent workflows in RS and EX on hard tasks (RS: 0.428 vs.\ 0.435/0.448; EX: 0.191 vs.\ 0.240/0.265). This gap pinpoints a decision bottleneck: the dominant failure mode is selecting \emph{which} tools to invoke and \emph{how} to interpret intermediate results, not tool execution itself. Advancing data agents thus demands stronger decision policies, not merely fewer execution errors.}

\noindent \bchange{\textbf{Takeaway 3: Thinking Models Plus Complex Architectures Can Backfire.}}
\bchange{For the same LLM, accuracy and cost vary substantially across workflow patterns (Table~\ref{tab:patterns_eff}), and selecting the right workflow can yield larger gains than switching to a higher-tier model. Notably, for explicit-reasoning (``thinking'') models, layering heavy external coordination atop internal deliberation introduces a double reasoning penalty, raising token cost with marginal quality gain (Table~\ref{tab:patterns_eff}). Design should match workflow complexity to each model's reasoning capacity rather than a uniform architecture.}

\section{CONCLUSION}

In this work, we propose FDABench, a comprehensive benchmark of 2,007 tasks spanning six data modalities with a multi-granularity evaluation framework that assesses both structured answers and open-ended reports for data agents over heterogeneous analytical scenarios. We design \textit{PUDDING}, an agentic framework that couples LLM-driven generation with iterative expert validation for scalable and reliable benchmark construction. Extensive experiments across 12 foundation models and four workflow patterns, covering general analytical agents, semantic operator frameworks, and RAG-based methods, reveal that agent performance is governed more by analytical topology and tool orchestration quality than by query complexity, and that pairing strong thinking models with complex workflows can even backfire.

\begin{acks}
This research is supported by Singapore MOE AcRF Tier-2 grant MOE-T2EP20223-0004. Wei Dong is supported by the National Research Foundation, Singapore, and the Cyber Security Agency of Singapore under the National Cybersecurity R\&D Programme and the CyberSG R\&D Programme Office (Award CRPO-GC3-NTU-001). Any opinions, findings, or recommendations expressed herein do not reflect the views of these agencies. We also thank Chuanjie Gong, Haoxuan Jia, Chuangxin Chu, Ruoxin Huang, Yichu Chen, Xiao He for their helpful feedback on this work.
\end{acks}

\bibliographystyle{ACM-Reference-Format}
\balance
\bibliography{ref}

\appendix
\section{Target System Descriptions}
\label{appendix:baselines}

This section provides detailed descriptions of the target systems evaluated in our experiments.

\subsection{General Analytical Query Systems}

\begin{itemize}[leftmargin=10.2pt]
\setlength{\itemsep}{0pt}
\setlength{\parsep}{0pt}
\setlength{\parskip}{0pt}

\item \textbf{Taiji~\cite{DBLP:journals/corr/taiji}:} A Model Context Protocol (MCP)-based data agent classified as \emph{Multi-agent workflow}. Through this architecture, Taiji supports multi-modal data analytics, dispatching modality-specific sub-plans to distributed servers.

\item \textbf{AOP~\cite{wang2025aop}:} A data agent system designed for automated pipeline orchestration in LLMs for answering complex queries, broadly classified as \emph{Planning Agent workflow}. It can also automatically and interactively optimize operators, which is similar to \emph{Reflection Agent workflow}.

\item \textbf{AgenticData~\cite{sun2025agenticdata}:} A data agent system classified as \emph{Multi-agent workflow}. AgenticData converts natural language queries into semantic plans through multi-agent collaboration for cross-domain analysis.

\item \textbf{MLE-STAR~\cite{mlestar}:} A data agent system classified as \emph{Multi-agent workflow}. MLE-Star leverages collaborative agents to translate natural language queries into executable analytical pipelines, enabling adaptive, cross-domain data reasoning.

\item \textbf{Teable~\cite{Teable}:} An open-source database agent classified as \emph{Planning Agent workflow}. Teable converts natural language prompts into executable SQL through offline planning strategies and provides analytical capabilities for structured databases.

\item \textbf{DeepAnalyze~\cite{deepanalyze}:} A fine-tuned foundation model with built-in tool orchestration designed for data analysis, classified as \emph{Tool-use Agent workflow}. DeepAnalyze integrates agentic behaviors directly into model weights, enabling autonomous invocation of code interpreters and analytical tools for analysis, visualization, and report generation.
\end{itemize}

\subsection{Semantic Operator Query Systems}

\begin{itemize}[leftmargin=10.2pt]
\setlength{\itemsep}{0pt}
\setlength{\parsep}{0pt}
\setlength{\parskip}{0pt}

\item \textbf{Palimpzest~\cite{palimpzestCIDR}:} A declarative framework optimizing LLM-driven analytics through semantic operators (\texttt{sem\_add\_columns}, \texttt{sem\_filter}, \texttt{retrieve}) for unified computation across structured and unstructured data.

\item \textbf{LOTUS~\cite{patel2024semanticoperators}:} A semantic operator programming model providing a unified DataFrame API with cross-modal operators (\texttt{sem\_sim\_join}, \texttt{sem\_agg}, \texttt{sem\_topk}) and batched inference for efficient computation.

\item \textbf{DocETL~\cite{shankar2024docetl}:} A framework optimizing complex document processing pipelines through LLMs, providing a declarative YAML-based interface for multi-modal processing workflows with specialized operators.
\end{itemize}

\subsection{RAG Systems}

\begin{itemize}[leftmargin=10.2pt]
\setlength{\itemsep}{0pt}
\setlength{\parsep}{0pt}
\setlength{\parskip}{0pt}

\item \textbf{CARROT~\cite{DBLP:journals/corr/CORAG}:} A cost-constrained RAG framework using Monte Carlo Tree Search for chunk combination order selection, considering correlations and non-monotonic utility within cost constraints.

\item \textbf{GraphRAG~\cite{GraphRAG}:} A graph-based method that uses LLMs to extract entities and relationships as nodes and edges, aggregates them into communities, and produces community summaries for contexts.

\item \textbf{HippoRAG2~\cite{DBLP:conf/nips/hipporag1, DBLP:journals/corr/hipporag2}:} A neurobiologically inspired graph-based RAG framework that enhances retrieval with integrated knowledge graphs and recognition memory.

\item \textbf{NaiveRAG~\cite{naiverag}:} A basic retrieval approach conducting vector similarity search for candidate chunks, followed by a reranker model.
\end{itemize}

\section{Dataset Generation}
\label{appendix:algorithm}

Algorithm~\ref{alg:hitl} summarizes the \textit{PUDDING} construction procedure, which executes in three phases: context grounding, agent-expert collaboration, and final validation/annotation.

\begin{algorithm}[t]
\caption{Dataset Generation Framework}
\label{alg:hitl}
\footnotesize
\SetKwInOut{Require}{Require}
\SetKwInOut{Ensure}{Ensure}
\Require{\textcolor{black}{\textcolor{black}{Seed database} instance $D_i \in \mathcal{D}$, original question $q$, \textcolor{black}{demonstration SQL} $s^*$}, max iterations $N_{max}$}
\Ensure{Benchmark task $\mathcal{P}$ with enhanced query $\tilde{q}$, gold answer $a^*$, frozen retrieved artifacts $\mathcal{I}$, Task DAG $\mathcal{G}$, and rubric $\mathcal{B}$}
\tcp{\textcolor{black}{Phase 1: Tree-Structured Context Grounding}}
$\mathcal{S} \leftarrow \textsc{SchemaExtract}(\textcolor{black}{D_i})$\;
$\mathcal{R} \leftarrow \textsc{ExecuteSQL}(\textcolor{black}{D_i}, \textcolor{black}{s^*})$ \tcp*{\textcolor{black}{Demonstration SQL for DB exploration}}
\textcolor{black}{$\mathcal{E} \leftarrow \textsc{EnterpriseKB}(q)$} \tcp*{\textcolor{black}{Retrieve demonstration cases}}
\textcolor{black}{$\sigma_0 \leftarrow \langle \mathcal{S}, \mathcal{R}, \mathcal{E} \rangle$} \tcp*{\textcolor{black}{Base State (Iter 0)}}
\textcolor{black}{$\mathcal{F} \leftarrow \{\sigma_0\}$, $\Pi \leftarrow \emptyset$, $t \leftarrow 0$} \tcp*{\textcolor{black}{Frontier, terminal paths}}
\While{\textcolor{black}{$\mathcal{F} \neq \emptyset$ \textbf{and} $t < N_{max}$}}{
    \textcolor{black}{$t \leftarrow t + 1$, $\mathcal{F}' \leftarrow \emptyset$}\;
    \ForEach{\textcolor{black}{$\sigma_v \in \mathcal{F}$}}{
        \textcolor{black}{$\mathcal{C}_v \leftarrow \textsc{SpawnCandidates}(\sigma_v, q)$} \tcp*{$b_v \geq 1$ branches}
        \ForEach{$(a, u) \in \mathcal{C}_v$}{
            $(o, p) \leftarrow \textsc{Exec}(a, u)$\;
            \textcolor{black}{$\sigma_{v'} \leftarrow \sigma_v \cup \{(a, u, o, p)\}$} \tcp*{Independent child context}
            \textcolor{black}{$d \leftarrow \textsc{SelfReflect}(\sigma_{v'})$}\;
            \lIf{\textcolor{black}{$d = \textsc{Continue}$}}{\textcolor{black}{$\mathcal{F}' \leftarrow \mathcal{F}' \cup \{\sigma_{v'}\}$}}
            \lElseIf{\textcolor{black}{$d = \textsc{Sufficient}$}}{\textcolor{black}{$\Pi \leftarrow \Pi \cup \{\textsc{Path}(\sigma_{v'})\}$}}
            \tcp{\textcolor{black}{\textsc{Prune}: discard branch}}
        }
    }
    \textcolor{black}{$\mathcal{F} \leftarrow \mathcal{F}'$}\;
}
\BlankLine

\tcp{Phase 2: Agent-Expert Collaboration}
\ForEach{\textcolor{black}{$\pi_i \in \Pi$}}{
Initialize: $\mathcal{H} \leftarrow \emptyset$, $k \leftarrow 0$, $\text{status} \leftarrow \textsc{PENDING}$\;
\textcolor{black}{$\mathcal{P}_0^{(i)} \leftarrow \textsc{Agent}(\pi_i, \sigma_{v_i}, q, s^*)$} \tcp*{Draft from each terminal path}
\textcolor{black}{$\mathcal{P}_{\text{current}} \leftarrow \mathcal{P}_0^{(i)}$}\;
\While{$k < N_{max}$ \textbf{and} $\text{status} \notin \{\textsc{ACCEPT}, \textsc{REJECT}\}$}{
    $k \leftarrow k + 1$\;
    \textcolor{black}{$\mathcal{P}_{\text{current}} \leftarrow \textsc{Reflect}(\mathcal{P}_{\text{current}})$} \tcp*{\textcolor{black}{Automated quality reflection}}
    $(\text{status}, \mathcal{C}_k) \leftarrow \textsc{ExpertReview}(\textcolor{black}{\mathcal{P}_{\text{current}}})$\;
    \If{$\text{status} = \textsc{REVISE}$}{
        $\mathcal{H} \leftarrow \mathcal{H} \cup \{(\textcolor{black}{\mathcal{P}_{\text{current}}}, \mathcal{C}_k)\}$\;
        \textcolor{black}{$\mathcal{P}_{\text{current}} \leftarrow \textsc{Agent}(\pi_i, \sigma_{v_i}, q, s^*, \mathcal{H})$}\;
    }
}
\BlankLine
\tcp{Phase 3: Finalization}
\If{$\text{status} = \textsc{REJECT}$ \textbf{or} $k = N_{max}$}{
    \textbf{continue} \tcp*{Skip this path}
}
\ElseIf{$\text{status} = \textsc{ACCEPT}$}{
    \If{\textsc{SingleSourceSolve}(\textcolor{black}{$\mathcal{P}_{\text{current}}$}, \textcolor{black}{$\sigma_{v_i}$})}{
        \textbf{continue}\;
    }
    $(\mathcal{G}, \mathcal{B}) \leftarrow \textcolor{black}{\textsc{Annotate}(\pi_i)}$\;
    \textcolor{black}{$\mathcal{P}_{\text{current}}$}.\text{dag} $\leftarrow \mathcal{G}$\;
    \textcolor{black}{$\mathcal{P}_{\text{current}}$}.\text{rubric} $\leftarrow \mathcal{B}$\;
    \Return \textcolor{black}{$\mathcal{P}_{\text{current}}$}\;
}
}
\end{algorithm}

\section{Task DAG Structure}
\label{appendix:dag}

For tasks requiring multi-step reasoning, \textit{PUDDING} derives a Task DAG (Directed Acyclic Graph) from each \textsc{Sufficient} terminal path's execution trace in Phase~1, formalizing tool invocation sequences and rationales into logical dependencies.

\begin{tcolorbox}[colback=gray!5, colframe=gray!50, title={DAG Node Types}, fonttitle=\bfseries\small]
\footnotesize
\begin{tabular}{@{}l@{\hspace{8pt}}p{0.6\linewidth}@{}}
\texttt{SQL\_QUERY} & Execute database queries, produce structured results \\
\texttt{RETRIEVE\_DOC} & Retrieve unstructured content (web, vector, file) \\
\texttt{EXTRACT\_EVIDENCE} & Extract key facts from retrieved documents \\
\texttt{COMPUTE} & Perform computation on SQL results \\
\texttt{VALIDATE} & Cross-validate with external knowledge \\
\texttt{SYNTHESIZE\_REPORT} & Generate final analytical output \\
\end{tabular}
\end{tcolorbox}

\begin{tcolorbox}[colback=gray!5, colframe=gray!50, title={DAG Edge Types}, fonttitle=\bfseries\small]
\footnotesize
\begin{tabular}{@{}l@{\hspace{8pt}}p{0.6\linewidth}@{}}
\texttt{HARD\_DEP} & Blocking dependency: node A must complete before B \\
\texttt{SOFT\_DEP} & Recommended but not required ordering \\
\texttt{ALT\_GROUP} & Alternative branches where $\geq k$ of $n$ must complete \\
\end{tabular}
\end{tcolorbox}

\noindent The DAG supports critical path identification for scheduling and parallel execution detection for nodes without mutual hard dependencies.

\section{Other Agent Benchmarks}
\label{appendix:other_benchmarks}

Recent agent benchmarks~\cite{DBLP:conf/iclr/ZhouX0ZLSCOBF0N24,DBLP:conf/iclr/0036YZXLL0DMYZ024,DBLP:conf/iclr/00020LCYPJ24} focus on capabilities for general agent tasks rather than data agent scenarios. GAIA~\cite{DBLP:conf/iclr/gaia} targets human-simple yet agent-challenging tasks, designing 466 questions across difficulty levels requiring understanding, reasoning, and tool-use proficiency. AgentBoard~\cite{DBLP:conf/nips/agentboard} provides multi-dimensional analytical evaluation with 1,013 examples across embodied AI, games, web, and tool use domains, assessing six agent capabilities including memory integration, planning decomposition, and world modeling. AgentBench~\cite{DBLP:conf/iclr/0036YZXLL0DMYZ024} evaluates cross-environment generalization across eight scenarios, including operating systems, databases, and knowledge graphs with multi-turn decision making. WebArena~\cite{DBLP:conf/iclr/ZhouX0ZLSCOBF0N24} focuses on web interaction scenarios, providing four fully-functional websites with 812 benchmark tasks for end-to-end web automation. MINT~\cite{DBLP:conf/iclr/00020LCYPJ24} targets sustained multi-turn problem-solving workflows. However, these benchmarks focus only on general capabilities rather than specialized skills for data preprocessing, statistical analysis, and data interpretation, lacking the capability to evaluate data agents' analytical performance over heterogeneous data.

\end{document}